\documentclass[accepted]{melba}
\usepackage{amsmath,amsfonts}
\usepackage{comment}
\usepackage{microtype}
\usepackage{float}

\melbaid{2025:002}  % This is provided upon by the publishing editor
\doi{https://10.59275/j.melba.2025-bea1}
\melbaauthors{LaBella and Calabrese}  % Note: this one is also used to set the pdf 'authors' metadata
\volume{3}
\firstpageno{38}  % Communicated by the publishing editor
\melbayear{2025}  % The publication year
\datesubmitted{05/2024}  % Date submitted to MELBA: mm/yyyy
\datepublished{03/2025}  % Today's date: mm/yyyy

\ShortHeadings{BraTS 2023 Meningioma Challenge Analysis}{LaBella and Calabrese}

\title{Analysis of the BraTS 2023 Intracranial Meningioma Segmentation Challenge}

\author{
    \name Dominic LaBella\aff{1}\orcid{0000-0003-1713-9538}
    \name Ujjwal Baid\aff{2,3}
    \name Omaditya Khanna\aff{4}
    \name Shan McBurney-Lin\aff{5}
    \name Ryan McLean\aff{6}
    \name Pierre Nedelec\aff{5}
    \name Arif Rashid\aff{7}
    \name Nourel Hoda Tahon\aff{8}
    \name Talissa Altes\aff{4}
    \name Radhika Bhalerao\aff{5}
    \name Yaseen Dhemesh\aff{8}
    \name Devon Godfrey\aff{1}
    \name Fathi Hilal\aff{8}
    \name Scott Floyd\aff{1}
    \name Anastasia Janas\aff{6}
    \name Anahita Fathi Kazerooni\aff{9,10}
    \name John Kirkpatrick\aff{1}
    \name Collin Kent\aff{1}
    \name Florian Kofler\aff{11,12,13,14}
    \name Kevin Leu\aff{15}
    \name Nazanin Maleki\aff{6}
    \name Bjoern Menze\aff{16,17}
    \name Maxence Pajot\aff{5}
    \name Zachary J. Reitman\aff{1}
    \name Jeffrey D. Rudie\aff{18,5}
    \name Rachit Saluja\aff{19}
    \name Yury Velichko\aff{20}
    \name Chunhao Wang\aff{1}
    \name Pranav Warman\aff{21}
    \name Maruf Adewole\aff{22}
    \name Jake Albrecht\aff{23}
    \name Udunna Anazodo\aff{24}
    \name Syed Muhammad Anwar\aff{25,26}
    \name Timothy Bergquist\aff{23}
    \name Sully Francis Chen\aff{21}
    \name Verena Chung\aff{23}
    \name Rong Chai\aff{23}
    \name Gian-Marco Conte\aff{27}
    \name Farouk Dako\aff{28}
    \name James Eddy\aff{23}
    \name Ivan Ezhov\aff{12,13}
    \name Nastaran Khalili\aff{9}
    \name Juan Eugenio Iglesias\aff{29,30,31}
    \name Zhifan Jiang\aff{25,26}
    \name Elaine Johanson\aff{32}
    \name Koen Van Leemput\aff{33}
    \name Hongwei Bran Li\aff{17,34,35}
    \name Marius George Linguraru\aff{25,26}
    \name Xinyang Liu\aff{25,26}
    \name Aria Mahtabfar\aff{4}
    \name Zeke Meier\aff{36}
    \name Ahmed W Moawad\aff{37}
    \name John Mongan\aff{5}
    \name Marie Piraud\aff{11}
    \name Russell Takeshi Shinohara\aff{10,38}
    \name Walter F. Wiggins\aff{39,40,41}
    \name Aly H. Abayazeed\aff{42}
    \name Rachel Akinola\aff{43}
    \name András Jakab\aff{44}
    \name Michel Bilello\aff{10,45}
    \name Maria  Correia de Verdier\aff{46}
    \name Priscila Crivellaro\aff{47}
    \name Christos Davatzikos\aff{10,45}
    \name Keyvan Farahani\aff{48}
    \name John Freymann\aff{48,49}
    \name Christopher Hess\aff{5}
    \name Raymond Huang\aff{50}
    \name Philipp Lohmann\aff{51,52}
    \name Mana Moassefi\aff{53}
    \name Matthew W. Pease\aff{54}
    \name Phillipp Vollmuth\aff{55,56}
    \name Nico  Sollmann\aff{57,58,59}
    \name David  Diffley\aff{60}
    \name Khanak K. Nandolia\aff{61}
    \name Daniel I Warren\aff{62}
    \name Ali Hussain\aff{63}
    \name Pascal Fehringer\aff{64}
    \name Yulia Bronstein\aff{65}
    \name Lisa Deptula\aff{66}
    \name Evan G. Stein\aff{67}
    \name Mahsa Taherzadeh\aff{68}
    \name Eduardo  Portela de Oliveira\aff{69}
    \name Aoife Haughey\aff{70}
    \name Marinos Kontzialis\aff{71}
    \name Luca Saba\aff{72}
    \name Benjamin Turner\aff{73}
    \name Melanie M. T. Brüßeler\aff{74}
    \name Shehbaz Ansari\aff{75}
    \name Athanasios Gkampenis\aff{76}
    \name David  Maximilian Weiss\aff{77}
    \name Aya  Mansour\aff{78}
    \name Islam H. Shawali\aff{79}
    \name Nikolay Yordanov\aff{80}
    \name Joel  M. Stein\aff{45}
    \name Roula Hourani\aff{81}
    \name Mohammed   Yahya Moshebah\aff{82}
    \name Ahmed Magdy Abouelatta\aff{83}
    \name Tanvir Rizvi\aff{84}
    \name Klara Willms\aff{6}
    \name Dann C. Martin\aff{85}
    \name Abdullah  Okar\aff{86}
    \name Gennaro  D’Anna\aff{87}
    \name Ahmed  Taha\aff{88}
    \name Yasaman  Sharifi\aff{89}
    \name Shahriar Faghani\aff{27}
    \name Dominic  Kite\aff{90}
    \name Marco  Pinho\aff{91}
    \name Muhammad  Ammar Haider\aff{92}
    \name Alejandro Aristizabal\aff{93,94}
    \name Alexandros Karargyris\aff{93}
    \name Hasan Kassem\aff{93}
    \name Sarthak Pati\aff{2,95,96}
    \name Micah  Sheller\aff{93,97}
    \name Michelle Alonso-Basanta\aff{7}
    \name Javier Villanueva-Meyer\aff{5}
    \name Andreas M Rauschecker\aff{5}
    \name Ayman Nada\aff{8}
    \name Mariam Aboian\aff{9}
    \name Adam E. Flanders\aff{98}
    \name Benedikt Wiestler\aff{17}
    \name Spyridon Bakas\aff{2,99,100,101}\orcid{0000-0001-8734-6482}
    \name Evan Calabrese\aff{5,41}
}
\affiliations{% <- trailing '%' to avoid unwanted indent
    \num 1 \addr Duke University Medical Center, Department of Radiation Oncology, Durham, NC, USA \\
    \num 2 \addr Division of Computational Pathology, Department of Pathology and Laboratory Medicine, Indiana University School of Medicine, Indianapolis, IN, USA \\
    \num 3 \addr Center for Federated Learning in Medicine, Indiana University, Indianapolis, IN, USA \\
    \num 4 \addr Department of Neurosurgery, Thomas Jefferson University, Philadelphia, PA, USA  \\
    \num 5 \addr University of California San Francisco, CA, USA \\
    \num 6 \addr Yale University, New Haven, CT, USA \\
    \num 7 \addr Department of Radiation Oncology, Perelman School of Medicine, University of Pennsylvania, Philadelphia, PA, USA \\
    \num 8 \addr University of Missouri, Columbia, MO, USA  \\
    \num 9 \addr Children’s Hospital of Philadelphia, University of Pennsylvania, Philadelphia, PA, USA \\
    \num 10 \addr Center for AI and Data Science for Integrated Diagnostics (AI2D) and Center for Biomedical Image Computing and Analytics (CBICA), University of Pennsylvania, Philadelphia, PA, USA \\
    \num 11 \addr Helmholtz AI, Helmholtz Munich, Germany \\
    \num 12 \addr Department of Informatics, Technical University Munich, Germany \\
    \num 13 \addr TranslaTUM - Central Institute for Translational Cancer Research, Technical University of Munich, Germany \\
    \num 14 \addr Department of Diagnostic and Interventional Neuroradiology, School of Medicine, Klinikum rechts der Isar, Technical University of Munich, Germany \\
    \num 15 \addr Center for Intelligent Imaging (ci2), Department of Radiology and Biomedical Imaging, University of California San Francisco (UCSF), San Francisco, CA, USA \\
    \num 16 \addr Biomedical Image Analysis and Machine Learning, Department of Quantitative Biomedicine, University of Zurich, Switzerland \\
    \num 17 \addr Department of Neuroradiology, Technical University of Munich, Munich, Germany \\
    \num 18 \addr University of San Diego, CA, USA \\
    \num 19 \addr Cornell University, Ithaca, NY, USA \\
    \num 20 \addr Department of Radiology, Northwestern University, Evanston, IL, USA \\
    \num 21 \addr Duke University Medical Center, School of Medicine, Durham, NC, USA \\
    \num 22 \addr Medical Artificial Intelligence (MAI) Lab, Crestview Radiology, Lagos, Nigeria \\
    \num 23 \addr Sage Bionetworks, USA \\
    \num 24 \addr Montreal Neurological Institute (MNI), McGill University, Montreal, QC, Canada \\
    \num 25 \addr Children's National Hospital, Washington DC, USA \\
    \num 26 \addr George Washington University, Washington DC, USA \\
    \num 27 \addr Mayo Clinic, Rochester, MN, USA \\
    \num 28 \addr Center for Global Health, Perelman School of Medicine, University of Pennsylvania, Philadelphia, PA, USA \\
    \num 29 \addr Athinoula A Martinos Center for Biomedical Imaging, Massachusetts General Hospital and Harvard Medical School, Boston, MA, USA \\
    \num 30 \addr Centre for Medical Image Computing, University College London, London, UK \\
    \num 31 \addr Computer Science and Artificial Intelligence Laboratory, Massachusetts Institute of Technology, Cambridge, MA, USA \\
    \num 32 \addr PrecisionFDA, U.S. Food and Drug Administration, Silver Spring, MD, USA \\
    \num 33 \addr Department of Applied Mathematics and Computer Science, Technical University of Denmark, Denmark \\
    \num 34 \addr Athinoula A Martinos Center for Biomedical Imaging, Massachusetts General Hospital, Boston, MA, USA \\
    \num 35 \addr University of Zurich, Switzerland \\
    \num 36 \addr Booz Allen Hamilton, McLean, VA, USA \\
    \num 37 \addr Mercy Catholic Medical Center, Darby, PA, USA \\
    \num 38 \addr Penn Statistics in Imaging and Visualization Center, Department of Biostatistics, Epidemiology, and Informatics \\
    \num 39 \addr Greensboro Radiology, Greensboro, NC, USA \\
    \num 40 \addr Radiology Partners, El Segundo, CA, USA \\
    \num 41 \addr Duke University Medical Center, Department of Radiology, Durham, NC, USA \\
    \num 42 \addr Neosoma Inc. Stanford Medicine, Stanford, CA, USA \\
    \num 43 \addr Lagos University Teaching Hospital, Lagos Nigeria. \\
    \num 44 \addr University of Zürich, Zürich,Switzerland \\
    \num 45 \addr Department of Radiology, Perelman School of Medicine, University of Pennsylvania, Philadelphia, PA, USA \\
    \num 46 \addr Department of Neuroradiology, Uppsala University, Sweden \\
    \num 47 \addr University of Toronto, Toronto, ON, Canada \\
    \num 48 \addr Cancer Imaging Program, National Cancer Institute, National Institutes of Health, Bethesda, MD, USA \\
    \num 49 \addr Leidos Biomedical Research, Inc., Frederick National Laboratory for Cancer Research, Frederick, MD, USA \\
    \num 50 \addr Department of Radiology, Brigham and Women's Hospital, Boston, MA, USA \\
    \num 51 \addr Institute of Neuroscience and Medicine (INM-4), Research Center Juelich, Juelich, Germany \\
    \num 52 \addr Department of Nuclear Medicine, University Hospital RWTH Aachen, Aachen, Germany \\
    \num 53 \addr Artificial Intelligence Lab, Department of Radiology, Mayo Clinic, Rochester, MN, USA \\
    \num 54 \addr Department of Neurosurgery, Indiana University, Indianapolis, IN, USA \\
    \num 55 \addr Department of Medical Image Computing, German Cancer Research Center (DKFZ), Heidelberg, Germany \\
    \num 56 \addr Department of Neuroradiology, University Hospital Bonn, Bonn Germany \\
    \num 57 \addr Department of Diagnostic and Interventional Radiology, University Hospital Ulm, Ulm, Germany \\
    \num 58 \addr Department of Diagnostic and Interventional Neuroradiology, School of Medicine, Klinikum rechts der Isar, Technical University of Munich, Munich, Germany \\
    \num 59 \addr TUM-Neuroimaging Center, Klinikum rechts der Isar, Technical University of Munich, Munich, Germany \\
    \num 60 \addr Fort Worth, TX, USA \\
    \num 61 \addr Department of Diagnostic and Interventional Radiology, All India Institute of Medical Sciences, Rishikesh, India \\
    \num 62 \addr Department of Neuroradiology, Washington University, St. Louis, MO, USA \\
    \num 63 \addr University of Rochester Medical Center, Rochester, NY, USA \\
    \num 64 \addr Faculty of Medicine, Jena University Hospital, Friedrich Schiller University Jena, Jena, Germany \\
    \num 65 \addr vRad (Radiology Partners), Minneapolis, MN, USA \\
    \num 66 \addr Ross University School of Medicine, Bridgetown, Barbados \\
    \num 67 \addr Department of Radiology, New York University Grossman School of Medicine, New York, NY, USA \\
    \num 68 \addr Department of Radiology, Arad Hospital, Tehran, Iran \\
    \num 69 \addr Department of Radiology, Faculty of Medicine, University of Ottawa, Cananda \\
    \num 70 \addr Department of Neuroradiology, JDMI, University of Toronto, TO, Canada \\
    \num 71 \addr Department of Radiology Northwestern University, Chicago, IL \\
    \num 72 \addr Department of Radiology, Azienda Ospedaliero Universitaria of Cagliari-Polo di Monserrato, Cagliari, Italy \\
    \num 73 \addr Department of Radiology, Leeds General Infirmary, Leeds, United Kingdom \\
    \num 74 \addr Ludwig Maximilians University, Munich, Bavaria, Germany \\
    \num 75 \addr Rush University Medical Center, Chicago, IL, USA \\
    \num 76 \addr Department of Neurosurgery, University Hospital of Ioannina, Ioannina, Greece \\
    \num 77 \addr Department of Neuroradiology, University Hospital Essen, Essen, North Rhine-Westphalia, Germany \\
    \num 78 \addr Egyptian Ministry of Health, Cairo, Egypt \\
    \num 79 \addr Department of Radiology, Kasr Alainy, Cairo University, Cairo, Egypt \\
    \num 80 \addr Faculty of Medicine, Medical University of Sofia, Sofia, Bulgaria \\
    \num 81 \addr Department of Radiology, American University of Beirut Medical center, Beirut, Lebanon \\
    \num 82 \addr Radiology Department, King Faisal Medical City, Abha, Saudi Arabia \\
    \num 83 \addr Department of Diagnostic and Interventional Radiology, Cairo University, Cairo, Egypt \\
    \num 84 \addr Department of Radiology and Medical Imaging, University of Virginia Health, Charlottesville, VA, USA \\
    \num 85 \addr Department of Radiology and Radiologic Sciences, Vanderbilt University Medical Center, TN, USA \\
    \num 86 \addr Faculty of Medicine, Hamburg University, Hamburg, Germany \\
    \num 87 \addr Neuroimaging Unit, ASST Ovest Milanese, Legnano, Milan, Italy \\
    \num 88 \addr University of Manitoba, Manitoba, Canada \\
    \num 89 \addr Department of Radiology, School of Medicine, Iran University of Medical Sciences, Iran, Tehran \\
    \num 90 \addr Department of Radiology, University Hospitals Bristol and Weston NHS Foundation Trust, Bristol, United Kingdom \\
    \num 91 \addr Department of Radiology, University of Texas Southwestern Medical Center, TX, USA \\
    \num 92 \addr CMH Lahore Medical College, Lahore, Pakistan \\
    \num 93 \addr MLCommons, Beaverton, OR, USA \\
    \num 94 \addr Factored AI, Palo Alto, CA, USA \\
    \num 95 \addr Center For Federated Learning in Medicine, Indiana University, Indianapolis, IN, USA \\
    \num 96 \addr Medical Working Group, MLCommons, San Francisco, CA, USA \\
    \num 97 \addr Intel Corporation, Santa Clara, CA, USA \\
    \num 98 \addr Department of Radiology, Thomas Jefferson University, Philadelphia, PA, USA \\
    \num 99 \addr Department of Radiology and Imaging Sciences, Indiana University School of Medicine, Indianapolis, IN, USA \\
    \num 100 \addr Department of Neurological Surgery, Indiana University School of Medicine, Indianapolis, IN, USA \\
    \num 101 \addr Department of Biostatistics and Health Data Science, Indiana University School of Medicine, Indianapolis, IN, USA 
}

\abstract{%   <- trailing '%' for backward compatibility of .sty file
   We describe the design and results from the BraTS 2023 Intracranial Meningioma Segmentation Challenge. The BraTS Meningioma Challenge differed from prior BraTS Glioma challenges in that it focused on meningiomas, which are typically benign extra-axial tumors with diverse radiologic and anatomical presentation and a propensity for multiplicity. Nine participating teams each developed deep-learning automated segmentation models using image data from the largest multi-institutional systematically expert annotated multilabel multi-sequence meningioma MRI dataset to date, which included 1000 training set cases, 141 validation set cases, and 283 hidden test set cases. Each case included T2, FLAIR, T1, and T1Gd brain MRI sequences with associated tumor compartment labels delineating enhancing tumor, non-enhancing tumor, and surrounding non-enhancing FLAIR hyperintensity. Participant automated segmentation models were evaluated and ranked based on a scoring system evaluating lesion-wise metrics including dice similarity coefficient (DSC) and 95\% Hausdorff Distance. The top ranked team had a lesion-wise median dice similarity coefficient (DSC) of 0.976, 0.976, and 0.964 for enhancing tumor, tumor core, and whole tumor, respectively and a corresponding average DSC of 0.899, 0.904, and 0.871, respectively. These results serve as state-of-the-art benchmarks for future pre-operative meningioma automated segmentation algorithms. Additionally, we found that 1286 of 1424 cases (90.3\%) had at least 1 compartment voxel abutting the edge of the skull-stripped image edge, which requires further investigation into optimal pre-processing face anonymization steps.
 }

\keywords{Meningioma, BraTS, Machine Learning, Segmentation, BraTS-Meningioma, Image Analysis Challenge, artificial intelligence, AI}

\begin{document}

% top matter
%\twocolumn[\maketitle]
% comment the preceedings and uncomment the following if the authors list + abstract is longer than one page
\maketitle
\twocolumn

% Introduction (or first section)
% \rule{\textwidth}{1pt}
\section{Introduction and Related Works}
	\enluminure{M}eningiomas are the most common primary brain tumors, and several common treatment modalities, including surgical resection and radiation therapy, require accurate delineation of tumor components \citep{ogasawara2021meningioma,rogers2017intermediate,rogers2020high}. When used clinically, meningioma MRI segmentation is often performed using T1-weighted (T1), T2-weighted (T2), T2-weighted-Fluid-Attenuated Inversion Recovery (FLAIR), and T1 post-contrast (T1Gd) multi-sequence brain magnetic resonance image (MRI) \citep{martz2022anocef}. Meningioma segmentation on brain MRI can be challenging due to the diverse morphology and location of meningiomas within the brain.  Anatomically, meningiomas arise from the arachnoid layer of the meninges between the dura mater and pia mater and commonly present at supratentorial sites of dural reflection, along the sphenoid sinus, and the skull base. Less commonly, meningiomas occur in intraventricular and suprasellar regions, the olfactory groove, and in the posterior fossa along the petrous bone \citep{labella2023asnr}. Examples of common anatomical locations of meningioma are depicted in Figure 1, which is an unmodified figure by Murek \citep{murek_localization_2024}. Their extra-axial location can frequently lead to their exclusion, in whole or in part, from brain MRI skull-stripping pre-processing steps. Radiographically, meningioma can present with a wide range of presentations which contributes to the difficulty in creating accurate generalizable meningioma automated segmentation models \citep{watts2014magnetic}. Commonly encountered radiographical variants and findings include en plaque meningioma, which is a plaque-like sessile extension of tumor along the meninges, cystic meningioma components, dural tail involvement extension, peri-tumoral edema, and numerous distinct lesions \citep{watts2014magnetic}.

    Recent advancements in segmentation techniques for brain tumors, particularly gliomas, through the application of deep learning and convolutional neural networks (CNNs), have shown promise in overcoming these challenges, offering increased accuracy and reproducibility compared to traditional methods \citep{pereira2016brain,havaei2017brain,bouget2022preoperative}. Since the inception of the Brain Tumor Segmentation (BraTS) challenge in 2012, they have been instrumental in propelling forward the field of brain tumor imaging segmentation by providing comprehensive datasets that facilitate the development and benchmarking of segmentation algorithms \citep{menze2014multimodal, bakas2017advancing}. The inaugural 2012 challenge had 35 training cases and 15 test cases focused solely on glioma, and the glioma dataset most recently increased to over 2000 cases included in the 2023 challenge. Studies by Menze et al. and Bakas et al. underscore the importance of the BraTS dataset in improving the segmentation accuracy for gliomas, leveraging multi-sequence MR images to improve the delineation of tumor tissues from non-tumorous brain matter \citep{menze2014multimodal,bakas2017advancing,gordillo2013state, icsin2016review}.
    
    In 2023, the BraTS organizing committee hosted new automated segmentation challenges to additionally focus on pediatric tumors, gliomas diagnosed in sub-Saharan Africa, brain metastasis, and meningioma \citep{brats_synapse,labella2023asnr,baid2021rsna,kazerooni2023brain,moawad2023brain}. Building on the foundation of prior BraTS challenges, the BraTS 2023 Intracranial Meningioma Segmentation Challenge aims to establish a community standard and benchmark for intracranial meningioma segmentation \citep{bakas2017advancing,labella2023asnr,calabrese2023brats}. We present a comprehensive analysis of segmentation performance across nine teams participating in the challenge focusing on key metrics: Enhancing tumor (ET) dice similarity coefficient (DSC), tumor core (TC) DSC, and whole tumor (WT) DSC, ET 95\% Hausdorff Distance (95HD), TC 95HD, and WT 95HD.  These metrics were evaluated on a lesion-wise basis to account for the possibility of multiple lesions. In many clinical scenarios, particularly in diseases such as meningioma where patients may present with multiple lesions of varying sizes, global metrics tend to average performance over the entire image volume. This averaging can mask suboptimal performance on smaller or less conspicuous lesions. By contrast, lesion‐wise evaluation allows each individual lesion to be assessed separately, thereby providing a more nuanced picture of an algorithm’s performance. For instance, a segmentation algorithm might achieve a high overall DSC by accurately segmenting larger lesions while missing or poorly delineating smaller ones. Evaluating the DSC and 95HD on a lesion-by-lesion basis highlights such discrepancies, which is particularly important for clinical decision-making where even a single missed lesion could be significant. Advantages and disadvantages of lesion-wise metrics are listed below.
    \paragraph{Advantages of lesion-wise metrics:}
    \begin{itemize}
    \item \textbf{Granular Assessment:} Evaluates each lesion individually, revealing performance variability hidden in global metrics.
    \item \textbf{Clinical Relevance:} Aligns with clinical needs by ensuring even small, critical lesions are accurately segmented.
    \item \textbf{Error Localization:} Identifies specific algorithm weaknesses on a per-lesion basis.
    \end{itemize}

    \paragraph{Disadvantages of lesion-wise metrics:}
    \begin{itemize}
    \item \textbf{Noise Sensitivity:} Small errors in tiny lesions can disproportionately impact metric values.
    \item \textbf{Definition Ambiguity:} Variability in defining individual lesions (especially confluent ones) may lead to inconsistent evaluations.
    \end{itemize}

    By evaluating each of the competing teams' automated segmentation algorithms' performance using lesion-wise metrics, we can identify state-of-the-art machine learning algorithm techniques. By doing so, we anticipate extension beyond the technical realm, to impacting patient outcomes, surgical approaches, radiation therapy planning, and understanding tumor behavior such as the propensity for an extra-axial location. As such, this study contributes to the technical field of medical imaging analysis and to the broader understanding of meningioma treatment and management strategies.

\begin{comment}
%%%%%%%%%%%%%%%%%%%%%%%%%%%%%%%%%%%%%%%%%%%%%%%%%%%%%%%%%%%%%%%%%%%%%%%%%%%
% Related works
%%%%%%%%%%%%%%%%%%%%%%%%%%%%%%%%%%%%%%%%%%%%%%%%%%%%%%%%%%%%%%%%%%%%%%%%%%%
% Make sure to put your work into context and include apporpriate citations.
% We do not have limits on citation counts.
\section{Related Works}
	Spatial alignment, or registration, between two images is a building block for estimation of deformable templates. Alignment usually involves two steps: a global affine transformation, and a deformable transformation (as in many optical flow applications).

	Use \verb| \citep{}| for reference that is part of the sentence, and \verb| \citep{}| for references in parenthesis. For example,  \citep{viola1997alignment} is awesome. Also, this is a citation~ \citep{viola1997alignment}.
\end{comment}

\section{Methods}
    \subsection{Challenge Data}
    
    BraTS Meningioma Challenge image data was contributed from 6 different United States academic medical centers: Duke University, Yale University, Thomas Jefferson University, University of California San Francisco, Missouri University, and University of Pennsylvania. Image data consisted of T1, T2, FLAIR, and T1Gd brain MRI sequences from patients with radiographic or pathologic diagnosis of intracranial meningioma. All data preprocessing was conducted using the FeTS tool, \citep{pati2022federated} and included conversion to NIfTI file format, co-registration, $1 \, \text{mm}^3$ isotropic resampling to the SRI24 atlas space, and automated skull stripping \citep{schwarz2019identification,thakur2019skull,thakur2020brain,juluru2020identification, smith2002fast}. 
    The skull stripping algorithm, part of the FeTS preprocessing workflow, was integral in removing non-brain tissue, including the skull and scalp, to isolate the intracranial structures. The brain extraction tool, widely used in neuroimaging pipelines, relies on deformable models and intensity-based thresholding to separate the brain from surrounding tissues \citep{smith2002fast}. In this challenge, skull stripping was critical for preserving patient anonymity by preventing potential face reconstruction from MRI data and to standardize data preparation across institutions. However, it should be noted that meningiomas often extend through the skull and skull-base foramina, and any extra-cranial portions of the tumors were implicitly excluded by this process \citep{smith2002fast}. Despite this limitation, skull stripping was applied to ensure consistency with other BraTS 2023 challenges and to minimize the inclusion of non-brain tissue.
    After image data pre-processing, tumor compartment labels were created using a comprehensive pre-segmentation, manual correction, and expert revision process, including enhancing tumor, non-enhancing tumor, and surrounding non-enhancing FLAIR hyperintensity (SNFH) as seen in Figure 2, which is an unmodified figure by LaBella et al. \citep{labella2023asnr, LaBella2024}. The initial pre-segmentation was performed using a deep convolutional neural network-based model \citep{isensee2021nnu}. Subsequently, 39 annotators manually reviewed and refined the segmentations. These annotators had varying levels of experience, from medical stu1dents to fellowship-trained neuroradiologists. Each annotator was given instructions on how to use the ITKSnap annotation software \citep{py06nimg} as well as a document on common errors of meningioma pre-segmentation as discussed in the dataset resource paper by LaBella et al \citep{LaBella2024}. After each annotator completed their manual corrections, the labels were sent to a final board-certified fellowship trained neuroradiologist (EC) for approval. This multi-step approach ensured the accuracy and consistency of the segmentation labels, incorporating multiple rounds of revision as needed to achieve high-quality final segmentations.  All participating institutions received Institutional Review Board and Data Transfer Agreement approvals before contributing data, ensuring compliance with relevant regulatory authorities.

    \begin{figure}[h]
              \centering
              % include first image
              \includegraphics[width=1\linewidth]{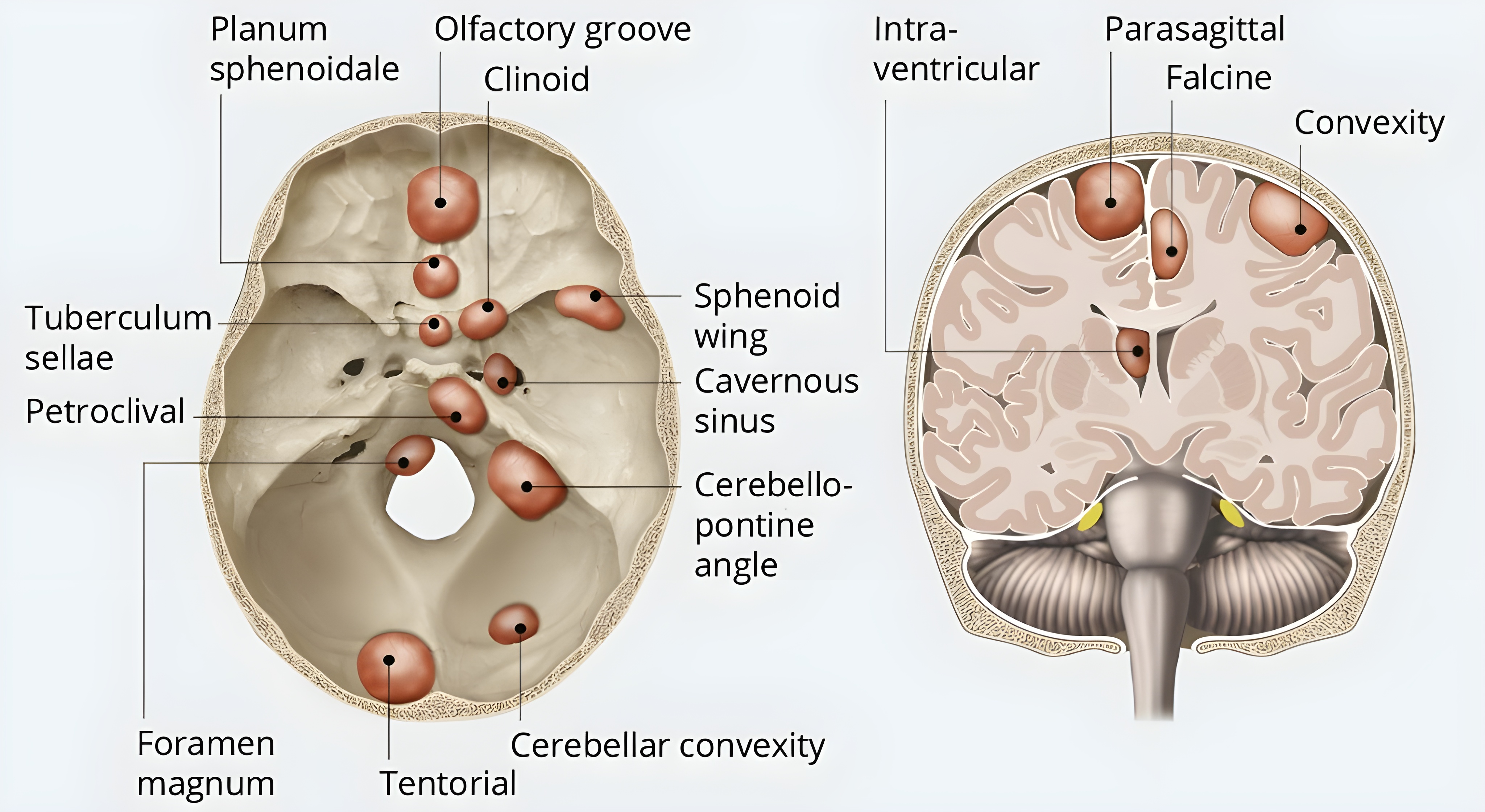}  
              \caption{Axial (left) and coronal (right) views of meningiomas at the most common locations in the skull. This is a modified figure as adapted from Murek under the CC-BY-4.0 license}
        \end{figure}

    \begin{figure}[h]
              \centering
              % include first image
              \includegraphics[width=1\linewidth]{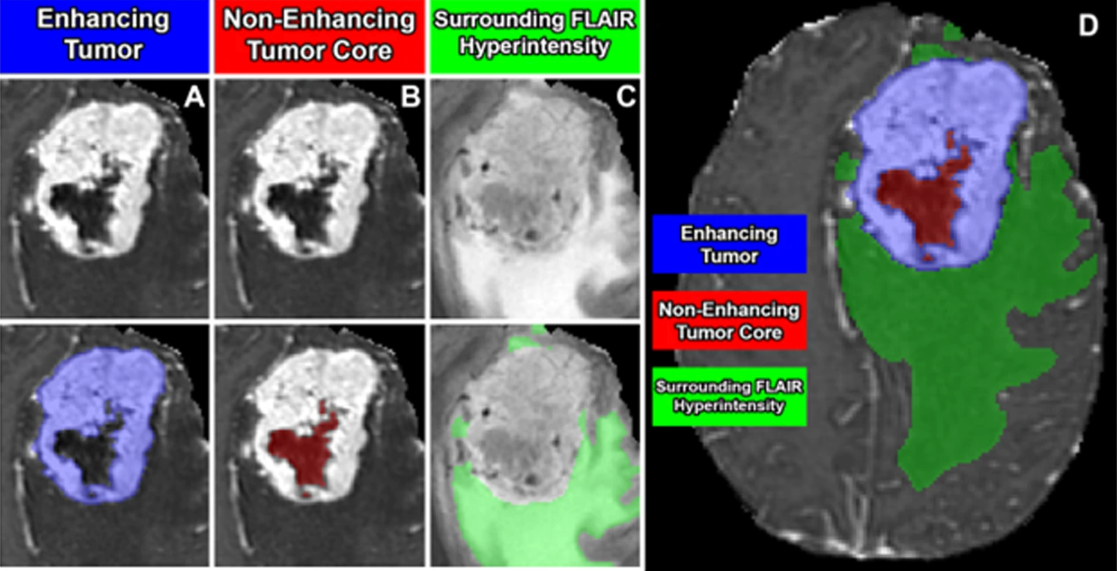}  
              \caption{Meningioma sub-compartments considered in the BraTS Pre-operative Meningioma Dataset. Image panels A-C denote the different tumor sub-compartments included in manual annotations; (A) enhancing tumor (blue) visible on a T1-weighted post-contrast image; (B) the non-enhancing tumor core (red) visible on a T1-weighted post-contrast image; (C) the surrounding FLAIR hyperintensity (green) visible on a FLAIR image; (D) combined segmentations generating the final tumor sub-compartment labels provided in the BraTS Pre-operative Meningioma Dataset.}
        \end{figure}

    \subsection{Challenge Procedures and Timeline}     
    
    The BraTS 2023 Intracranial Meningioma Segmentation Challenge was hosted on the Synapse platform using the BraTS Pre-operative Meningioma Dataset \citep{brats_synapse}. To access the challenge dataset and to be eligible for submission of automated segmentation models, participants were required to register as a participating team on the Synapse platform. Registered teams developed automated segmentation algorithms that trained on multi-sequence MRI of pre-treatment intracranial meningioma with associated ground truth labels that were released to the participating teams in May 2023. 
    
    In June 2023, each of the participating teams had access to additional validation data consisting of multi-sequence MRI cases. For validation data, teams were able to assess segmentation performance of their models by submitting predicted labels through the Synapse platform, but individual ground truth segmentations were not made publicly available. From July until August 2023, participating teams utilized the validation dataset to fine tune their segmentation models and compose short paper manuscripts.  At the end of the validation phase, each participating team uploaded their optimal automated segmentation model and respective manuscript as an MLCube container, which was used for evaluation in the testing phase.  During the testing phase, the BraTS organizing committee internally evaluated each of the participating team’s automated segmentation models on the hidden test set of pre-operative meningioma cases with ground truth labels. 
    
    \subsection{Algorithm Evaluation}  
    
    During the testing phase, the BraTS organizing committee evaluated metrics on three regions of interest including ET, TC, and WT. ET was solely the enhancing tumor compartment label. TC was the combination of enhancing tumor and non-enhancing tumor compartment labels. WT was the combination of enhancing tumor, non-enhancing tumor, and SNFH compartment labels. Note that the term “whole tumor” was used across the BraTS 2023 cluster of challenges for consistency; however, this term is not entirely accurate for meningioma, where SNFH typically does not contain any tumor but rather represents associated vasogenic edema.
    Metrics used for evaluation included DSC and the 95HD and were evaluated on a lesion-wise level. The DSC is a measure used to quantify the similarity between two samples, which, in this context, refers to the overlap between automated segmentation and the expert annotated ground truth labels for each respective tumor compartment. The 95HD observed in the segmentation results. The 95HD was used in lieu of the standard 100\% Hausdorff Distance to account for smaller lesions that may suffer from overestimates of the standard 100\% Hausdorff Distance. For previous BraTS challenges, a global DSC was used for challenge rankings. However, lesion-wise metrics were adopted for the 2023 challenge as there was greater potential for multiple distinct lesions in a single patient image (most notably for the metastasis and meningioma sub-challenges). Distinct lesions were identified by performing a 1 voxel symmetric dilation on the ground truth WT masks, and then evaluating a 26-connectivity 3D connected component analysis to determine if overlap between distinct lesions exists \citep{rudie2023brats}. A case’s lesion-wise DSC and 95HD scores are calculated based on equations (1) and (2) respectively, where L is the number of ground truth lesions and (true positive (TP) + false negative (FN)) is equal to L \citep{saluja2023lesion}. A predicted lesion is counted as a TP if at least 1 predicted voxel overlaps with the respective ground truth's respective region of interest mask. A lesion is counted as a FN if the model does not predict any voxels within the ground truth's respective region of interest mask. A predicted lesion is counted as a false positive (FP) if the model predicts a distinct lesion that does not overlap with any ground truth lesions' voxels. The lesion-wise scoring system assigned a specific lesion’s region of interest a DSC score of 0 and a 95HD score of 374 for FP or FN. These equations effectively calculate the average DSC or 95HD values across all of the predicted lesions for a given case. The scoring system also excluded evaluation for ground truth lesions smaller than 50 voxels to avoid evaluation of false ground truth lesions missed in dataset review. This threshold was discussed and decided by fellowship trained neuroradiologists after ground truth label review \citep{saluja2023lesion,rudie2023brats}. Evaluation of submissions was performed on  MLCommons' MedPerf, an open federated AI/ML evaluation platform \citep{karargyris2023federated}. MedPerf automated the evaluating pipeline by running the participants' models on the testing datasets of each contributing site’s data and calculating evaluation metrics on the resulting predictions. Finally, the Synapse platform retrieved the metrics results from the MedPerf server and ranked them to determine the winner \citep{MLCube2024, pati2023gandlf, karargyris2023federated}.
    
    \begin{equation}
    \text{Lesion-wise Dice Score} = \frac{\sum_{i}^{L} \text{Dice}(I_i)}{\text{TP} + \text{FN} + \text{FP}}
    \end{equation}
    \begin{equation}
    \text{Lesion-wise 95HD} = \frac{\sum_{i}^{L} \text{HD}_{95}(I_i)}{\text{TP} + \text{FN} + \text{FP}}
    \end{equation}
    
    \subsection{Participant Ranking and Workshop Proceedings}  
    
    The BraTS organizing committee internally evaluated each of the participating team’s automated segmentation models on the hidden test set of pre-operative meningioma cases to determine lesion-wise metrics for both DSC and 95HD for each of the three regions of interest. The participants were ranked against each other for each region of interest's lesion-wise metric independently. A total of 6 independent rankings were calculated to reflect the two metrics, DSC and 95HD, for each of the ET, TC, and WT regions of interest. Then a “BraTS segmentation score” was calculated based on the average of each independent lesion-wise region of interest metric rankings. For example, if a team had the 3rd best ET DSC, 2nd best TC DSC, 3rd best WT DSC, 3rd best ET 95HD, 2nd best TC 95HD, and 4th best WT 95HD, then that team would have an average ranking of (3+2+3+3+2+4) / 6 = 2.83 as their BraTS segmentation score.  The BraTS segmentation score was used to determine the final participant rankings relative to one another. The three top-ranked teams were invited to present their findings at the BraTS workshop at the 2023 MICCAI Annual Meeting held in Vancouver, Canada; although final rank was hidden until the workshop. At the BraTS workshop, the BraTS organizing committee announced the final placement of the three top-ranked teams. Monetary awards of \$1400, \$1000, and \$800 were presented to the three top-ranked teams, respectively.
    
    \subsection{Challenge Results Analysis} 
    
    Overall participant and individual team statistical analysis of DSC and 95HD lesion-wise performance was performed using Python and Microsoft Excel (Excel). Analysis included calculation of participant average, standard deviation, and median DSC and 95HD for each region of interest; overall average and median DSC and 95HD across all participants for each region of interest; volume calculations of each lesion; and number of abutting voxels of lesions compared to the pre-processed brain MRI.
    
    \subsection{Analysis of Tumor Abutment of Brain Masks}
    
    Given the extra-axial location of meningiomas, we sought to evaluate the proportion of meningiomas that were potentially cropped or excluded by the automated skull stripping process. To determine the volume of each compartment label and the number of tumor compartment voxels that were directly abutting the edge of the skull-stripped images for each case, the NumberOfEdgeNeighbors.py script was internally run by BraTS organizers, which evaluated each of the 1483 meningioma MRI cases \citep{dlabella29_meningiomaanalysis_2024}. This analysis was performed internally due to restricted access to the hidden test dataset. To determine significance of association of WT volume compared to abutting voxels, the Pearson Correlation Coefficient was calculated with associated p-value with a significance level of 0.05.

\section{Results}
    A total of 1000 training (70\%) multi-sequence pre-operative meningioma MRI cases, 141 validation cases (10\%), and 283 test cases (20\%) were utilized within the BraTS Meningioma Challenge (Table 1) in adherence with standard machine learning protocols. 
    
    A total of 9 participating teams submitted automated segmentation models to MLCube for the BraTS Challenge 2023: Intracranial Meningioma. The statistical summary of the teams' performances is outlined in Tables 2 and 3; which list the calculated DSCs and 95HD, respectively. The maximum recorded average DSC for ET, TC, and WT are 0.899, 0.904, and 0.871, respectively; and the minimum recorded average 95HD for ET, TC, and WT are 23.9, 21.8, and 31.4, respectively; highlighting the upper bounds of team performance within the challenge. The overall challenge summary statistics across all participating teams are listed in Table 4 for both DSC and 95HD. Figure 3 shows violin plots of DSC and 95HD scores for the ET, TC, and WT regions across all of the participating teams. Figure 4 shows a comparison of the predictions for a single testing set case from each of the top 3 participant's algorithms.
    
    \begin{table}[h]
    \centering
    \caption{This table presents the total number of cases in each of the training, validation, and testing sets. Note that the training data was released with ground truth labels. Note that the validation data was released without ground truth labels. The included institutions are Duke University (Duke), Thomas Jefferson University (TJU), Missouri (Miss), University of Pennsylvania (Penn),  University of California San Francisco (UCSF), and Yale University (Yale).}
    \begin{tabular}{@{}lccc@{}}
    \toprule
    & \textbf{Train} & \textbf{Validation} & \textbf{Test} \\ 
    \midrule
    \textbf{Total Count}    & 1000           & 141                 & 283           \\
    Duke         & 315            & 46                  & 91            \\
    TJU & 236      & 34                  & 68            \\
    Miss                & 132            & 16                  & 33            \\
    Penn & 31         & 4                   & 9             \\
    UCSF & 126 & 18              & 35            \\
    Yale       & 160            & 23                  & 47            \\
    Release Date & May 2023& June 2023 & Never released\\
    \bottomrule
    \end{tabular}
    \end{table}

    \begin{table*}[h]
    \centering
    \caption{Team DSC scores, average $\pm$ SD (median), for ET, TC, and WT regions of interest. Combined team rankings for each respective metric.}
    \begin{tabular}{@{}lcccc@{}}
    \toprule
    \textbf{Team Name} & \textbf{ET DSC} & \textbf{TC DSC} & \textbf{WT DSC} & \textbf{Rank (ET, TC, WT)} \\
    \midrule
    NVAUTO          & 0.899 $\pm$ 0.189 (0.976) & 0.904 $\pm$ 0.180 (0.976) & 0.871 $\pm$ 0.198 (0.964) & 1, 1, 1 \\
    CNMC\_PMI2023   & 0.876 $\pm$ 0.217 (0.968) & 0.867 $\pm$ 0.227 (0.968) & 0.851 $\pm$ 0.231 (0.953) & 2, 3, 2 \\
    blackbean      & 0.870 $\pm$ 0.222 (0.969) & 0.879 $\pm$ 0.206 (0.969) & 0.845 $\pm$ 0.226 (0.957) & 3, 2, 3 \\
    Sherlock        & 0.854 $\pm$ 0.234 (0.958) & 0.850 $\pm$ 0.239 (0.959) & 0.831 $\pm$ 0.244 (0.945) & 4, 4, 4 \\
    huilin          & 0.830 $\pm$ 0.276 (0.959) & 0.820 $\pm$ 0.258 (0.958) & 0.761 $\pm$ 0.297 (0.897) & 5, 5, 6 \\
    i\_sahajmistry  & 0.799 $\pm$ 0.291 (0.954) & 0.773 $\pm$ 0.303 (0.949) & 0.764 $\pm$ 0.296 (0.932) & 6, 7, 5 \\
    Kurtlab-UW      & 0.790 $\pm$ 0.237 (0.896) & 0.774 $\pm$ 0.250 (0.892) & 0.745 $\pm$ 0.257 (0.872) & 7, 6, 8 \\
    MIA             & 0.775 $\pm$ 0.305 (0.940) & 0.757 $\pm$ 0.307 (0.941) & 0.751 $\pm$ 0.306 (0.916) & 8, 8, 7 \\
    UMNiverse       & 0.007 $\pm$ 0.084 (0.000) & 0.027 $\pm$ 0.078 (0.006) & 0.241 $\pm$ 0.290 (0.092) & 9, 9, 9 \\
    \bottomrule
    \end{tabular}
    \end{table*}

    \begin{table*}[h]
    \centering
    \caption{Team 95\% Hausdorff distances, average $\pm$ SD (median), for ET, TC, and WT regions of interest. Combined team rankings for each respective metric.}
    \begin{tabular}{@{}lcccc@{}}
    \toprule
    \textbf{Team Name} & \textbf{ET 95HD} & \textbf{TC 95HD} & \textbf{WT 95HD} & \textbf{Rank (ET, TC, WT)} \\
    \midrule
    NVAUTO         & 23.9 $\pm$ 68.5 (0.96)  & 21.8 $\pm$ 64.6 (1.0)   & 31.4 $\pm$ 71.8 (1.0)   & 1, 1, 1 \\
    CNMC\_PMI2023  & 30.0 $\pm$ 80.9 (1.0)   & 31.7 $\pm$ 83.5 (1.0)   & 35.2 $\pm$ 86.8 (1.62)  & 2, 3, 2 \\
    blackbean     & 34.3 $\pm$ 82.0 (1.0)    & 29.9 $\pm$ 75.0 (1.0)   & 41.2 $\pm$ 84.3 (1.0)    & 3, 2, 4 \\
    Sherlock      & 34.3 $\pm$ 87.5 (1.07)   & 35.1 $\pm$ 88.3 (1.93)  & 39.7 $\pm$ 93.1 (1.94)  & 4, 4, 3 \\
    Kurtlab-UW    & 39.9 $\pm$ 90.2 (2.0)    & 45.9 $\pm$ 95.8 (2.0)    & 56.0 $\pm$ 100.4 (1.05) & 5, 5, 6 \\
    huilin        & 46.9 $\pm$ 104.2 (1.0)   & 47.7 $\pm$ 105.2 (1.41) & 55.9 $\pm$ 106.9 (3.61) & 6, 6, 5 \\
    i\_sahajmistry & 56.5 $\pm$ 108.9 (1.0)  & 64.1 $\pm$ 112.5 (1.41) & 66.2 $\pm$ 111.0 (2.0)  & 7, 8, 8 \\
    MIA           & 61.4 $\pm$ 117.6 (1.41) & 61.6 $\pm$ 118.0 (1.73) & 64.2 $\pm$ 119.5 (2.83) & 8, 7, 7 \\
    UMNiverse     & 371.4 $\pm$ 314.9 (374.0) & 150.1 $\pm$ 151.5 (81.5) & 158.8 $\pm$ 146.9 (189.1) & 9, 9, 9 \\
    \bottomrule
    \end{tabular}
    \end{table*}

    \begin{figure*}[htbp]
        \centering
        \includegraphics[width=\linewidth]{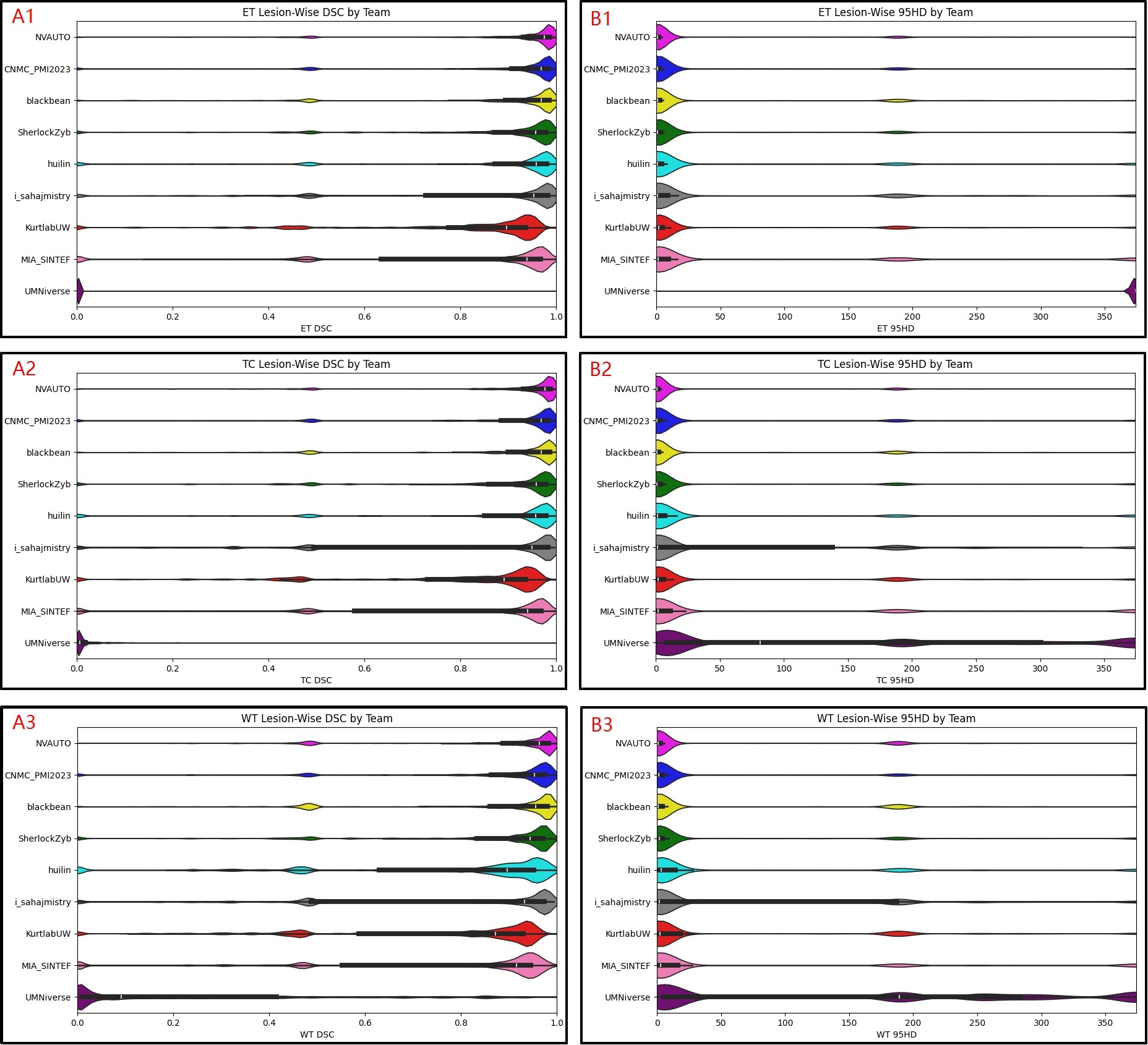}
        \caption{Violin plots of DSC and 95HD scores for the ET, TC, and WT regions across all of the participating teams. The subplots are organized as: A1 (ET DSC), A2 (TC DSC), A3 (WT DSC), B1 (ET 95HD), B2 (95HD TC), B3 (95HD WT).}
    \end{figure*}
    
    \begin{figure*}[htbp]
        \centering
        \includegraphics[width=\linewidth]{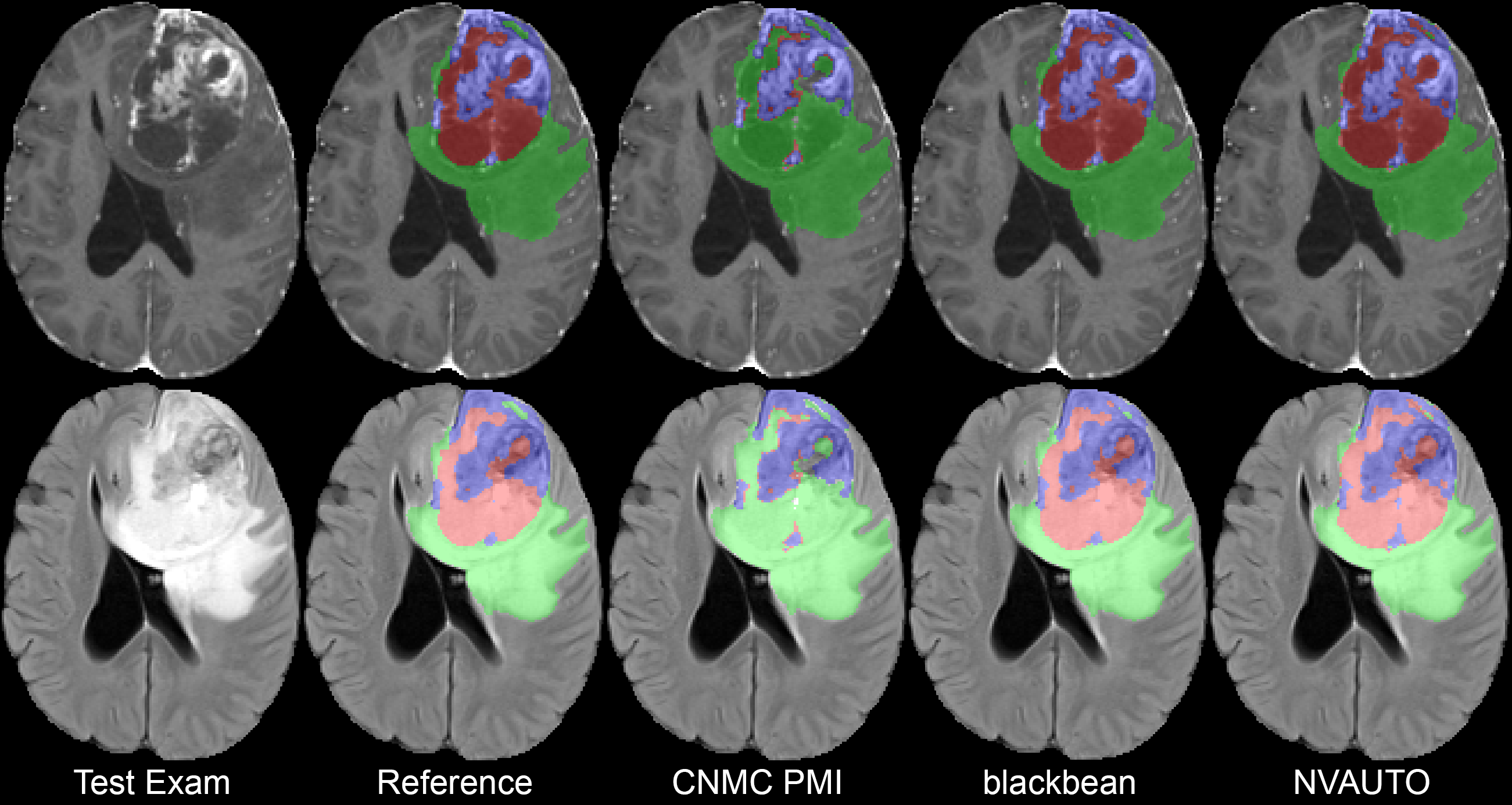}
        \caption{Image panels demonstrating the different predictions of the top 3 teams for a pre-operative meningioma testing set case as seen on T1Gd (top row) and FLAIR (bottom row) MRI.}.
    \end{figure*}

    \begin{table*}[h]
    \centering
    \caption{Summary statistics for DSC and 95\% Hausdorff Distance (95HD) across ET, TC, and WT regions of interest for 9 participating teams in a segmentation task. The DSC (Dice Similarity Coefficient) and 95HD metrics are presented with their respective statistics: Average, Std (Standard Deviation), Median, (Q1, Q3) (1st and 3rd Quartiles), and either Max or Min values as applicable.}
    \label{tab:combined_summary_statistics}
    \begin{tabular}{@{}lcccccc@{}}
    \toprule
    \textbf{Statistic} & \textbf{ET DSC} & \textbf{TC DSC} & \textbf{WT DSC} & \textbf{ET 95HD} & \textbf{TC 95HD} & \textbf{WT 95HD} \\
    \midrule
    \textbf{Average}       & 0.830 & 0.820 & 0.764 & 39.9  & 46.0  & 55.9  \\
    \textbf{Std}           & 0.234 & 0.239 & 0.257 & 87.5  & 95.76 & 100.4 \\
    \textbf{Median}        & 0.958 & 0.958 & 0.933 & 1.00  & 1.414 & 2.00  \\
    \textbf{(Q1, Q3)}      & [0.872, 0.981] & [0.850, 0.979] & [0.630, 0.972] & [1.00, 4.24] & [1.00, 6.27] & [1.00, 14.5] \\
    \textbf{Max Avg } & 0.899 & 0.904 & 0.871 & 23.9  & 21.8 & 31.4     \\
    \textbf{Max Med } & 0.976 & 0.976 & 0.964 & 1.00  & 1.00  & 1.00   \\
    
    \bottomrule
    \end{tabular}
    \end{table*}

    The top 3 ranked teams for the BraTS Meningioma Challenge included NVAUTO, blackbean, and CNMC\_PMI2023 \citep{capellan2023brats,myronenko2023auto3dseg,huang2023exploring}. Each of these teams were invited to give an oral presentation of their findings and methods at MICCAI 2023 in Vancouver, Canada. Their lesion-wise DSC and 95HD and median lesion-wise DSC, averaged over ET, TC, and WT are listed in Table 5. NVAUTO's winning MONAI Auto3DSeg framework and blackbean's STU-Net framework are open-source and freely accessible (Apache License 2.0) \citep{myronenko2023auto3dseg, myronenko2018mri, monai2020project, huang2023exploring, huang2023stu}. CNMC\_PMI2023 utilized an ensemble of nnUNet, an open-source and freely accessible Apache License 2.0 model, and SWIN-transformer, a freely accessible MIT License model \citep{capellan2023brats}.  The team leaders, submitted short paper titles, and technical aspects of their algorithms are listed below:

    \begin{enumerate}
        \item \textbf{NVAUTO}: Andriy Myronenko et al., \textit{Auto3DSeg for Brain Tumor Segmentation from 3D MRI in BraTS 2023 Challenge} \citep{myronenko2023auto3dseg}.\\
        \\
        NVAUTO employed the Auto3DSeg tool from MONAI for brain tumor segmentation using 3D MRI scans \citep{myronenko2018mri, monai2020project, myronenko2023auto3dseg}. The core of the model architecture used in the challenge was SegResNet, a U-Net based convolutional neural network designed for semantic segmentation tasks. This model utilizes an encoder-decoder structure, incorporating repeated ResNet blocks with batch normalization and deep supervision, which helps guide training through multiple layers of the network \citep{myronenko2018mri}. To improve performance and robustness, several data augmentation techniques were applied, including random affine transformations, flipping, intensity scaling, shifting, noise addition, and blurring. These augmentations help the model generalize better by simulating variations in MRI data. The loss function used for training combined Dice loss and focal loss, with the goal of addressing class imbalance by emphasizing harder-to-segment areas and penalizing inaccurate segmentations of minority classes. Additionally, the loss was summed across deep-supervision sublevels, meaning the network computed losses at various resolution scales to refine the segmentation. The optimizer employed was AdamW, with an initial learning rate of $2 \, \text{e}^{-4}$, gradually reduced to zero using the cosine annealing scheduler. This adaptive optimization method, combined with a learning rate decay, ensures better convergence and prevents overfitting. Weight decay regularization of $1 \, \text{e}^{-5}$ was also used to prevent overfitting by penalizing large weights in the model. The Auto3DSeg framework was designed to be user-friendly, requiring minimal manual input. It automates several stages of the training and optimization process, making it accessible even to non-experts. Advanced users can fine-tune various parameters, such as hyperparameters and model architecture, for improved performance. The training setup leveraged 8 NVIDIA V100 GPUs with 16 GB of memory each, and a 5-fold cross-validation process was used to ensure generalizability across different MRI datasets, further improving the model's accuracy and robustness.
        
        \item \textbf{blackbean}: Ziyan Huang et al., \textit{Evaluating STU-Net for Brain Tumor Segmentation} \citep{huang2023exploring}. \\
        \\
        Blackbean utilized the Scalable and Transferable U-Net (STU-Net) model in the 2023 BraTS Challenge \citep{huang2023exploring, huang2023stu}. STU-Net builds upon the nnU-Net architecture but introduces key modifications to enhance its scalability and transferability for large-scale medical image segmentation tasks \citep{huang2023stu, isensee2021nnu}. The model’s architecture ranges from 14 million to 1.4 billion parameters, enabling flexibility depending on computational resources. The core innovation lies in the incorporation of residual connections to address gradient diffusion and downsampling blocks within each encoder stage for more efficient feature extraction. STU-Net also utilizes nearest-neighbor interpolation with a 1×1×1 convolution layer for upsampling, which improves the model's ability to generalize and transfer learning across different imaging tasks \citep{huang2023stu}. A compound scaling strategy ensures balanced growth of both encoder and decoder components, optimizing both depth and width. Pre-trained on the TotalSegmentator dataset, which covers 104 foreground classes, the model demonstrates robust transferability to the BraTS brain tumor segmentation task \citep{wasserthal2022totalsegmentator}. Blackbean adhered to the default data pre-processing, data augmentation, and training procedures provided by nnU-Net, and they utilized the SGD optimizer with a Nesterov momentum of 0.99 and a weight decay of $1 \, \text{e}^{-3}$ \citep{isensee2021nnu}. The batch size was fixed at 2, and each epoch consisted of 250 iterations. The learning rate decay followed the poly learning rate policy: ${(1 - epoch/1000)}^{0.9}$.
        Data augmentation techniques used during training included additive brightness, gamma, rotation, scaling, mirror, and elastic deformation. The pre-training patch size on the TotalSegmentator dataset was 128 × 128 × 128 \citep{wasserthal2022totalsegmentator}. Fine-tuning patch sizes on downstream tasks were automatically configured by nnU-Net \citep{isensee2021nnu}.
              
        \item \textbf{CNMC\_PMI2023}: Daniel Capellán-Martín et al., \textit{Model Ensemble for Brain Tumor Segmentation in Magnetic Resonance Imaging} \citep{capellan2023brats}.  \\
        \\
        CNMC\_PMI2023 used an ensemble-based approach. The ensemble strategy combines two state-of-the-art deep learning models: nnU-Net and Swin UNETR \citep{isensee2021nnu, tang2022self}. The 3D nnU-Net model was trained using five-fold cross-validation, with input images divided into patches of 128 x 160 x 112 \citep{isensee2021nnu}. The output consisted of three channels corresponding to the three tumor sub-regions. A combined Dice loss and cross-entropy loss function was employed, optimized using the stochastic gradient descent (SGD) algorithm with Nesterov momentum (learning rate: 0.01, momentum: 0.99, weight decay: $3 \, \text{e}^{-5}$). Inference was conducted using a sliding window approach.  The vision transformer-based 3D Swin UNETR model was trained using five-fold cross-validation, with input patches of 96 x 96 x 96 voxels \citep{tang2022self}. The output was four channels: three tumor sub-regions and background. The combined Dice loss and focal loss function was optimized using the AdamW optimizer (learning rate: 0.0001, momentum: 0.99, weight decay: $3 \, \text{e}^{-5}$).
        To improve segmentation accuracy and robustness, predictions from nnU-Net and Swin UNETR were ensembled. The ensembling process involved combining outputs for the tumor regions (WT, TC, and ET) from each model across the five cross-validation folds. 
        Given the task’s emphasis on lesion-wise evaluation, a post-processing step was developed to clean small disconnected regions <50 voxels. Training for nnU-Net models was conducted on an NVIDIA A100 GPU with 40GB of memory, while Swin UNETR models were trained on NVIDIA A5000 (24GB) and A6000 (48GB) GPUs. Hyperparameter optimization was carried out using the Optuna framework \citep{optuna_2019}.

    \end{enumerate}
    
    \begin{table}[h]
        \centering
        \caption{This table shows the lesion-wise metrics for each of the top 3 ranked teams in the BraTS Challenge 2023: Intracranial Meningioma challenge. Note that a precision of 4 decimals is used due to the close final results amongst top participating teams.}
        \begin{tabular}{lccc}
            \toprule
             & \textbf{Average} & \textbf{Median} & \textbf{Average} \cr
             \textbf{Team Name} & \textbf{DSC} & \textbf{DSC} & \textbf{95HD} \cr
            \midrule
            NVAUTO & 0.8909 & 0.9855 & 25.70 \cr
            Blackbean & 0.8643 & 0.9861 & 34.84 \cr
            CNMC\_PMI2023 & 0.8638 & 0.9855 & 32.30 \cr
            \bottomrule
        \end{tabular}
    \end{table}

    Of note, NVAUTO placed in the top 2 for each of the five distinct BraTS 2023 automated segmentation challenges. NVAUTO came in first place for Meningioma, BraTS-Africa, Brain Metastases; and came in second place for Adult Glioma and Pediatric Tumors. For the meningioma challenge, NVAUTO had a total of 228 of 283 testing phase cases with a tumor core DSC $\geq$ 0.90. Additionally, during the public validation phase, as described in their in-person oral presentation at MICCAI, NVAUTO reported achieving an average composite DSC of 0.935, which is substantially higher than their testing phase top score of 0.891, which suggests some degree of overfitting \citep{myronenko2023auto3dseg}.
    
    \subsection{Notable Challenge Cases and Statistics}
    The top median DSC across all participants for a specific case was a perfect score of 1.00 for enhancing tumor and tumor core as shown in Figure 5. Note that there was a total of 33 enhancing tumor voxels, and 23 were abutting the edge of the MR image.  The cases with the next highest overall median and average DSC are shown in Figure 6. Note that the ET volumes are qualitatively much greater in these particularly high DSC cases.
    
    There were two meningioma cases with somewhat unusual imaging appearance as shown in Figures 7 and 8 which had poor performance for test performance metrics across all participants.  Notably, they had a majority of non-enhancing tumor making up the TC and WT. These lesions correspond to heavily calcified lesions with little or no visible enhancement and low signal intensity on all provided sequences.
    
    \begin{figure}[h]
              \centering
              % include first image
              \includegraphics[width=1\linewidth]{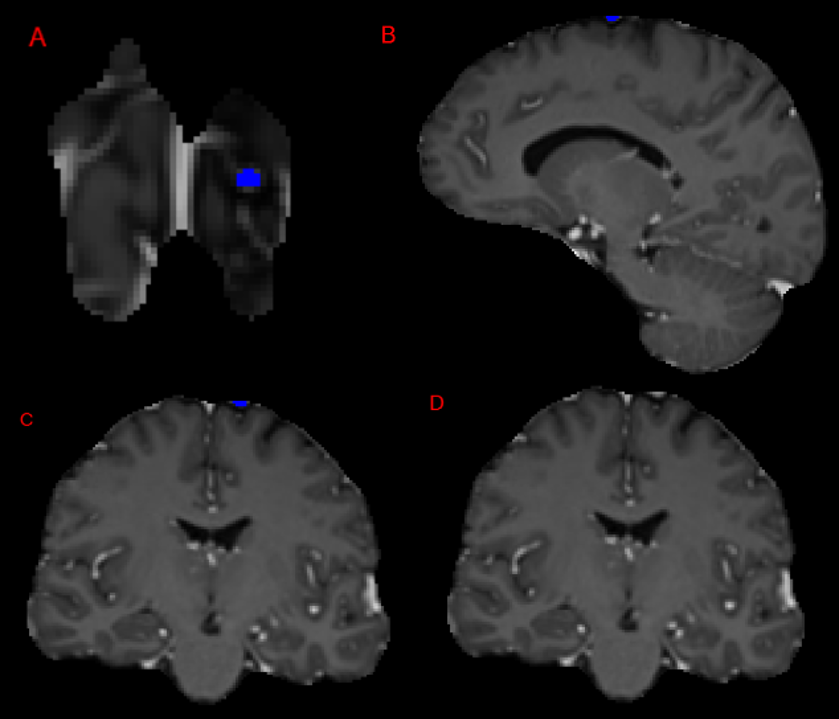}  
              \caption{Image panels of the top scored individual test case with a median participant DSC of 1.00, 1.00, and 1.00 and average participant DSC of 0.993, 0.881, and 0.882 for enhancing tumor, tumor core, and whole tumor; respectively. Tumor ground truth sub-compartment labels annotated on axial (A), sagittal (B), and coronal (C) views of a T1Gd MRI head case. Panel D depicts the tumor abutting the edge of the skull-stripped brain without a label.}
        \end{figure}
    
    \begin{figure}[h]
              \centering
              % include first image
              \includegraphics[width=1\linewidth]{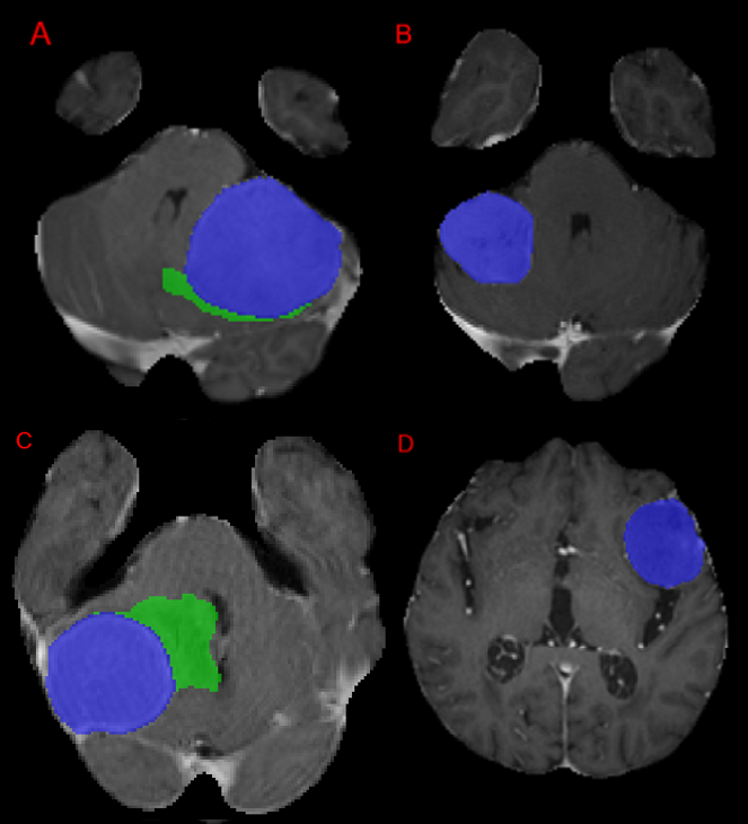}  
              \caption{Image panels that depict participants median enhancing tumor and tumor core DSC of (A) 0.993, (B) 0.991, (C) 0.991, and (D) 0.991 (0.990 for TC). Note that the averages were 0.876, 0.878, 0.864, 0.875 respectively, which is due to a single team having a DSC of approximately 0.002 for each case, whereas every other team had an ET DSC above 0.95 for each case.}
        \end{figure}
    
    \begin{figure}[h]
              \centering
              % include first image
              \includegraphics[width=1\linewidth]{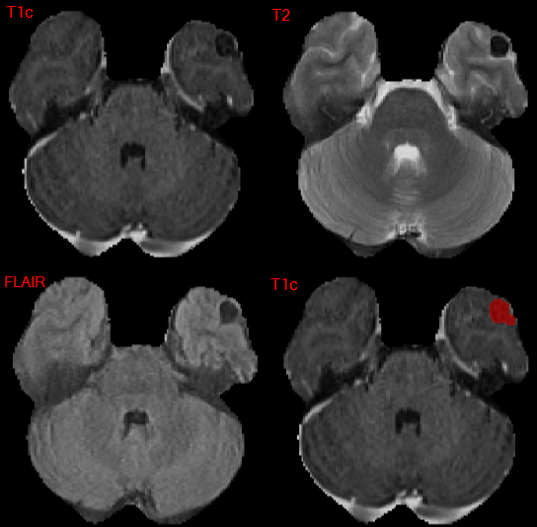}  
              \caption{Image panels depicting a completely calcified extra-axial non-enhancing meningioma that had a median participant DSC of 1.00, 0.00, and 0.00 and average participant DSC of 0.888, 0.175, and 0.174 for enhancing tumor, tumor core, and whole tumor; respectively.}
        \end{figure}
    
    \begin{figure}[h]
              \centering
              % include first image
              \includegraphics[width=1\linewidth]{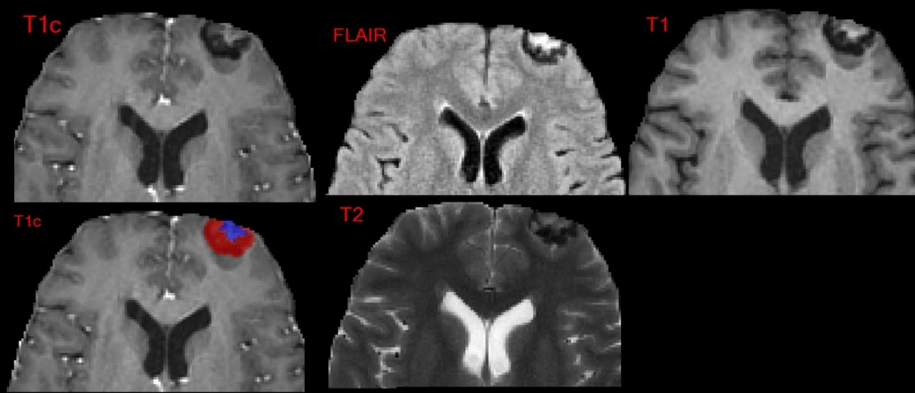}  
              \caption{MRI study demonstrating the worst performing meningioma case for NVAUTO with an ET DSC of 0.00, TC DSC of 0.00, and WT DSC of 0.338. The blue and red labels represent ground truth enhancing tumor and non-enhancing tumor, respectively, which combined make up the tumor core region of interest.}
        \end{figure}
    
    \begin{figure}[h]
              \centering
              % include first image
              \includegraphics[width=1\linewidth]{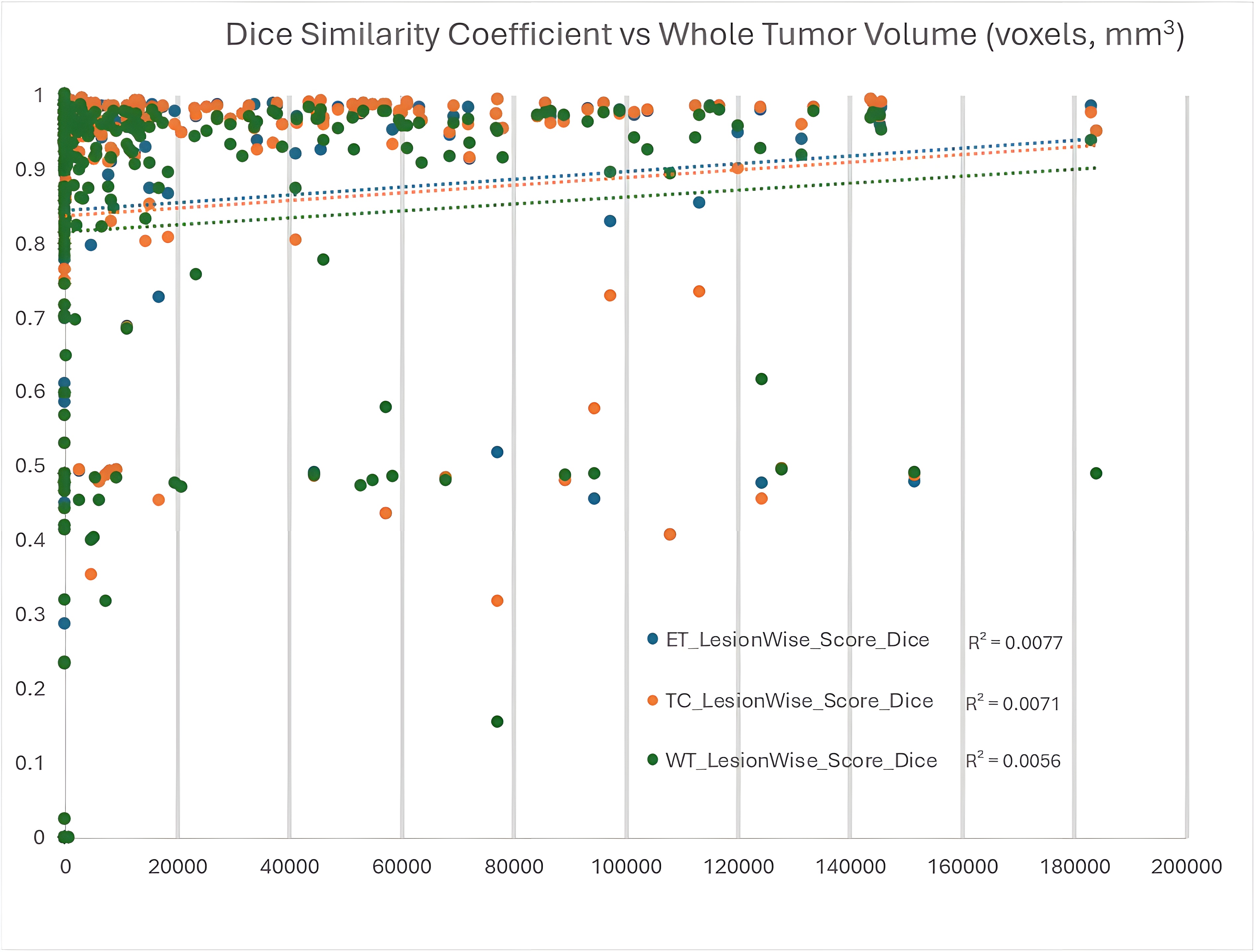}  
              \caption{Scatterplot graph depicting the overall participant median DSC for each region of interest to the WT volume for each of the test cases.}
        \end{figure}

    Each region of interest’s DSC was compared to WT volume for each respective case as shown in Figure 8.  There was a nonsignificant positive linear correlation between DSC vs WT volume for each of the three regions of interest, ET, TC, and WT; with p values of 0.696, 0.689, and 0.741, respectively. There was a nonsignificant logarithmic correlation between DSC vs WT volume for each of the three regions of interest, ET, TC, and WT; with p values of 0.102, 0.093, and 0.200, respectively (not shown in figure). Notably, Figure 9 also demonstrates a significant number of cases with a lesion-wise DSC of approximately 0.5 for each of the regions of interest.  This is due to the lesion-wise metrics penalizing false positives with a value of 0 for the respective prediction and false negatives with a value of 0 for missed ground truth lesions; combined with a very strong performance for another ground truth lesion. 
    
    Skull-stripping resulted in 908 of 1000 training cases, 129 of 141 validation cases, and 257 of 283 test cases’ meningioma labels (1286 of 1424, 90.3\% overall) having at least 1 compartment voxel abutting the edge of the skull-stripped image edge as seen in the example case in Figure 10. Of the 257 aforementioned cases, the average and median number of abutting voxels was 628.7 and 394 voxels respectively. Figure 11 shows the relationship between the number of abutting voxels and the WT volume ($\text{R}^2$ = 0.190 and p = 0.002).
    
    \begin{figure}[h]
              \centering
              % include first image
              \includegraphics[width=1\linewidth]{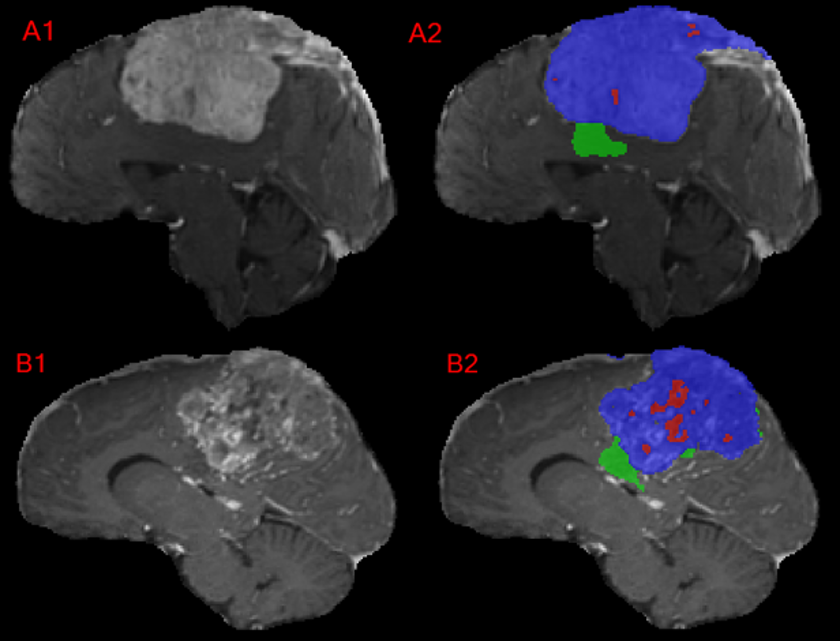}  
              \caption{Image panels with the tumor sub-regions annotated on sagittal views of T1Gd MRI head cases. Panels A1 and A2 represent a meningioma tumor with 4514 voxels abutting the skull-stripped boundary. Panels B1 and B2 represent a meningioma tumor with 3911 voxels abutting the skull-stripped boundary.}
        \end{figure}
    
    \begin{figure}[h]
              \centering
              % include first image
              \includegraphics[width=1\linewidth]{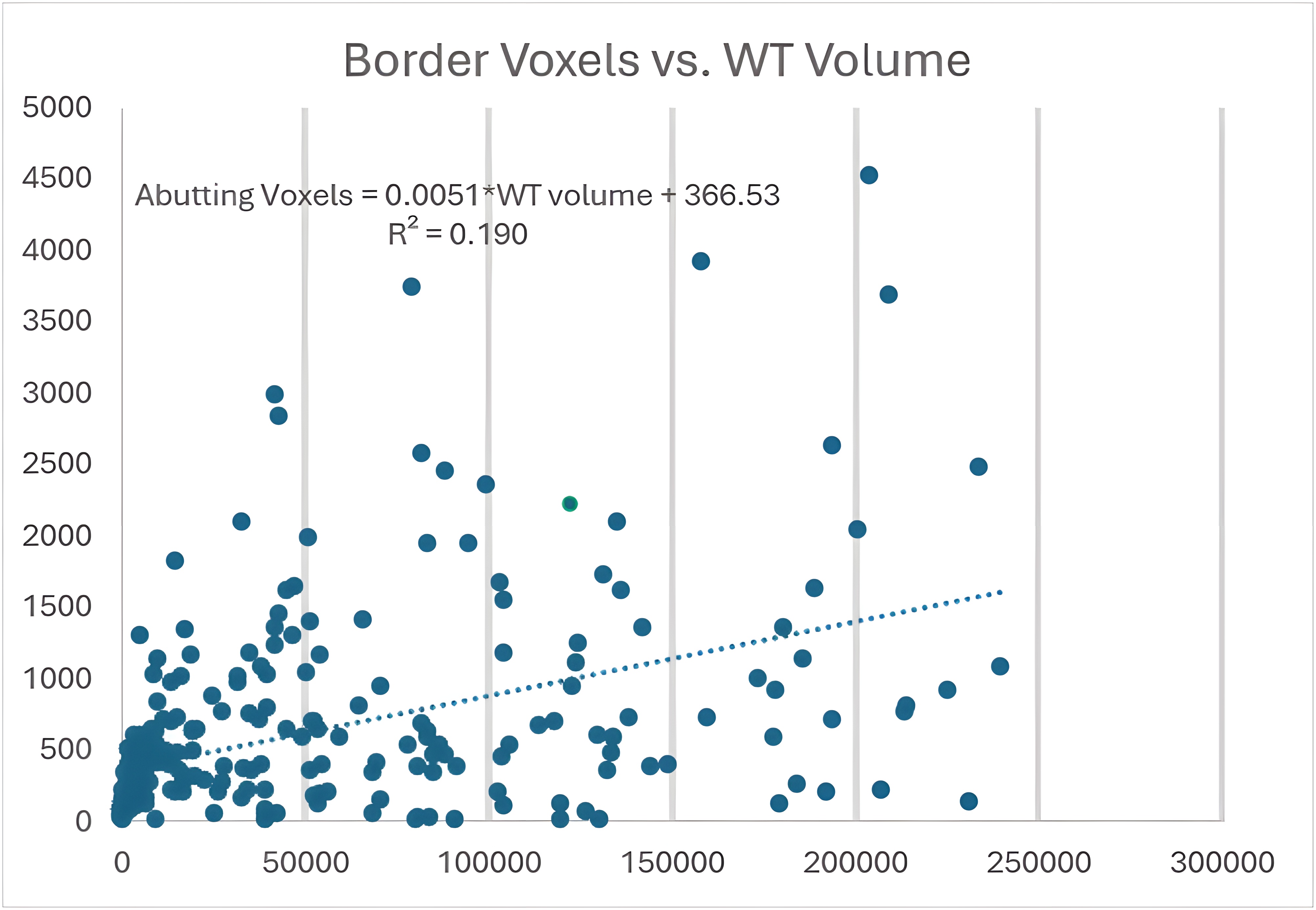}  
              \caption{Scatterplot depicting the relationship between WT size and the number of abutting voxels.}
        \end{figure}

    While global metrics (i.e. those used in previous BraTS challenges) were not used for ranking of the 2024 challenge, we have included Figure 12 to show the relationship between the complete tumor volume and each of the global metrics in cases with only a single lesion for each of the regions of interest. Note that there is a noticeable trend towards improved global DSC, global 95HD, and global sensitivity as complete volume increases.\\
    
    \begin{figure*}[htbp]
              \centering
              % include first image
              \includegraphics[width=1\linewidth]{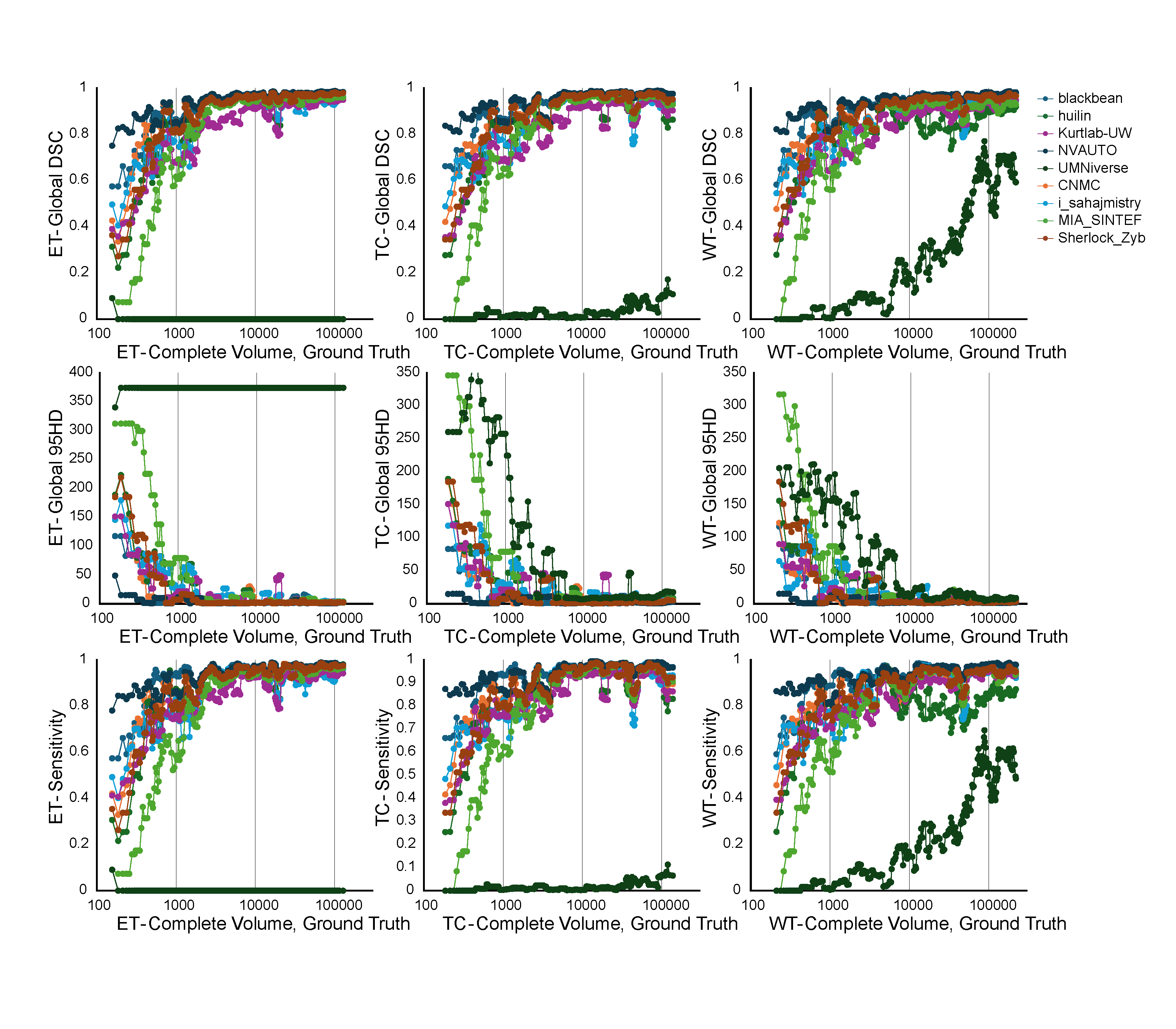}  
              \caption{Plots of sliding window average global DSC, global 95HD, and sensitivity for each of the ET, TC, and WT regions of interest for the subset of BraTS testing cases that only had 1 ground truth lesion. The number of subjects with a single lesion is ET (N=255), TC (N=255), and WT (N=252), for each respective label.}
        \end{figure*}

\section{Discussion}
    \subsection{Challenge Summary}
    The BraTS 2023 Intracranial Meningioma Segmentation Challenge provided unprecedented insight into the state-of-the-art performance in pre-operative meningioma segmentation, leveraging the largest multi-institutional systematically expert annotated multilabel meningioma MRI dataset to date \citep{calabrese2023brats, LaBella2024}. The challenge saw remarkable performances, particularly from the NVAUTO team, in DSC and 95HD across ET, TC, and WT region of interest segmentation tasks. It was notable that DSC scores were, on average, higher for the meningioma challenge compared to all other BraTS 2023 segmentation sub-challenges, which may be due to the relatively lower complexity of meningiomas compared to other tumor types and/or the relatively high quality and consistency of the meningioma dataset. These lesion-wise DSC and 95HD scores should be considered as benchmarks for future pre-operative meningioma segmentation evaluation. Note that lesion-wise metrics are essential for segmentation tasks with potential for more than one distinct lesion as global metrics may still remain relatively high even if a single, smaller lesion is completely missed by the segmentation algorithm.
    
    \subsection{The Best Algorithm and Caveats}
    The superior performance of NVAUTO, with lesion-wise median DSC values of 0.976, 0.976, and 0.964 lesion-wise average DSC values of 0.899, 0.904, and 0.871, for ET, TC, and WT respectively, signifies a notable advancement in automated meningioma segmentation algorithms. These results not only demonstrate the feasibility of achieving high accuracy in meningioma segmentation, but also suggest that deep learning models can effectively adapt to the diverse morphology and anatomical locations of meningiomas.
    NVAUTO’s base algorithm, known as AutoSeg3D, an open-source and Pytorch-based framework, is particularly adaptable to a variety of medical imaging automated segmentation challenges \citep{myronenko2023auto3dseg}. Auto3DSeg allows for auto-scaling to available GPUs; 5-fold training with SegResNet, DiNTS, and SwinUNETR models; and performing inference and ensembling using each of the multiple trained models.  For the challenges, NVAUTO found the SegResNet model to be the most accurate and was ultimately the optimal model architecture selected by Auto3DSeg for each of the automated segmentation challenges \citep{myronenko2023auto3dseg}. Additionally, Auto3DSeg, allows for use of a variety of image modalities and is not limited to just MRI. Auto3DSeg is advertised to run with a GPU RAM $\geq$ 8GB. However, for the 2023 BraTS challenges, NVAUTO used 8 x 16GB NVIDIA V100 machines \citep{myronenko2023auto3dseg}. Future studies should assess the ability to use Auto3DSeg on more widely available, consumer grade GPUs and non-NVIDIA branded GPUs.
    
    \subsection{Overall Segmentation Performance}
    The challenge results revealed a broad range of performances among the 9 participating teams, with DSC scores for the ET ranging from 0.899 to 0.007. Such variability underscores the challenge's complexity and the diverse approaches to segmentation. When comparing NVAUTO’s meningioma segmentation performance compared to their performance on other challenges, it was found that they performed the best on meningioma’s DSC for ET and TC regions of interest but performed only second best for DSC for WT as shown in Table 6. NVAUTO’s algorithm showed that Sub-Saharan Africa gliomas had a higher DSC for WT than all other tumor types.  Since meningioma tend to have a smaller overall whole tumor volume and a higher ET or TC to WT ratio compared to glioblastoma, it can be hypothesized that there is less available training information to create as accurate SNFH compartment labels; and thereafter, the WT region of interest \citep{ogasawara2021meningioma, gilard2021diagnosis, khandwala2021many}.  
    
    \begin{table}[h]
    \centering
    \caption{Average DSC for ET, TC, and WT for the top performing team, NVAUTO, using their algorithm, Auto3DSeg, across each of the different BraTS 2023 segmentation challenges. Note that 3 decimal precision is shown for meningioma to emphasize the small increase in DSC for non-ET compared to ET.}
    \label{tab:dsc_scores}
    \begin{tabular}{@{}lccc@{}}
    \toprule
    & \textbf{ET} & \textbf{TC} & \textbf{WT} \\
    & \textbf{(Avg DSC)} & \textbf{(Avg DSC)} & \textbf{(Avg DSC)} \\
    \midrule
    Meningioma          & 0.899 & 0.904 & 0.870 \cr
    Glioma              & 0.810 & 0.830 & 0.840 \cr
    
    Sub-Saharan  & 0.790 & 0.840 & 0.910 \cr
    Africa &&& \cr
    Metastasis          & 0.600 & 0.650 & 0.620 \cr
    Pediatric           & 0.550 & 0.780 & 0.840 \cr
    \bottomrule
    \end{tabular}
    \end{table}
    
    \subsection{Limitations of the BraTS Meningioma Benchmark}
    Our analysis revealed that all participating teams performed relatively poorly on heavily calcified meningiomas with little or no enhancing tumor. For example, teams had an average TC DSC of only 0.156 for one particular non-enhancing meningioma case. This relatively poor performance is presumably related to the relative rarity of this imaging appearance of meningioma, and the fact that only a small number of such cases were included in the training dataset. Future datasets should include more cases of exclusively non-enhancing tumor to allow for more generalizable automated segmentation models. It is important to consider that radiotherapy plans only consider a single GTV represented by the TC, and whether these indeterminate TC regions are labeled as non-enhancing vs enhancing regions would not impact the resulting treatment volumes \citep{rogers2017intermediate, rogers2020high}.
    
    Note that in Figure 9, the lines of best fit for ET, TC, and WT trend towards improved DSC with increased WT volume. In an automated segmentation challenge, this could cause a higher perceived test set overall DSC score if a larger proportion of larger tumors are included in the test set compared to the overall population. Therefore, it is important to ensure balance within the training, validation, and test sets regarding tumor size, which was not explicitly done for this iteration of the challenge.

    Another notable limitation of our study is the absence of explicit testing on out-of-distribution (OOD) cases. While our model demonstrated strong performance on the provided BraTS 2023 meningioma dataset, all data were derived from a limited number of institutions, and no evaluation was performed on data from external sources or significantly different MRI acquisition protocols. Consequently, the generalization ability of the model to cases from different populations, MRI machines, or acquisition settings remains untested. Future work should focus on assessing the model’s performance on OOD data to ensure its robustness and applicability in real-world clinical environments. Techniques such as domain adaptation and cross-institutional validation will be crucial to improve the model’s generalization capabilities and reliability across diverse clinical settings.
    
    \subsection{Future Directions}
    Future studies involving meningioma automated segmentation should focus on the most important clinical volumes, particularly the tumor core which comprises the radiotherapy gross tumor volume (GTV). Additionally, future studies should focus on the segmentation of meningioma along the intra-axial and extra-axial border with various face anonymization pre-processing techniques, due to the high frequency for meningioma to be excluded by skull-stripping as demonstrated by our results \citep{watts2014magnetic, schwarz2019identification, schwarz2021changing}.
    
    This study performed an analysis of the propensity of meningioma to abut the skull-stripped image, thereby having the potential to exclude portions of the meningioma within the skull-stripped image. Due to skull-stripping resulting in 1286 of 1483 meningioma having at least 1 compartment voxel abutting the boundary of the skull-stripped image, future studies should evaluate a different pre-processing anonymization technique that will allow for inclusion of more volume of intracranial meningioma. Schwarz et al. describes a promising mri\_reface technique that performs face anonymization by modifying the MR head image face and ear regions to represent an average human face and ear, thereby preserving the vast majority of the MR head image \citep{schwarz2019identification,schwarz2021changing}. Bischoff et al. describes mri\_deface, a defacing tool that removes facial features by assigns a probability of voxel being “face” or “brain” and removes voxels that have non-zero probability of being “face” but zero probability of being “brain” \citep{theyers2021multisite, bischoff2007technique}.

    Furthermore, while current segmentation approaches predominantly utilize MRI, atypical meningiomas—such as those that are completely non-enhancing or heavily calcified—remain challenging to automate segmentation due to low signal intensity and limited contrast. In these cases, integrating additional imaging modalities could prove highly beneficial. PET imaging, especially with tracers like $^{68}\mathrm{Ga}$-DOTATATE, provides metabolic information that can help differentiate viable tumor tissue from calcified or fibrotic areas \citep{prasad202268ga}. Likewise, CT offers superior spatial resolution for delineating calcifications and osseous involvement, which is critical when meningiomas extend into or invade bone \citep{salah2019can}. Indeed, studies have demonstrated that PET and CT can outperform MRI in identifying bony invasion and calcification in meningiomas \citep{galldiks2023advances, salah2019can}. Thus, the incorporation of multi-modal imaging may lead to more robust segmentation algorithms capable of addressing the full spectrum of meningioma presentations.
    
    RTOG 0539, a phase II trial of observation for low-risk meningiomas and of radiotherapy for intermediate- and high-risk meningiomas, describes radiation treatment planning and target volume protocols that should be used for meningiomas \citep{rogers2017intermediate, rogers2020high}. For radiation planning, they only required use of pre-operative and post-operative contrast-enhanced MRI. RTOG 0539 defines the GTV to encompass the tumor bed on postoperative-enhanced MRI and to include any residual nodular enhancement. They also state that trailing linear dural tail and cerebral edema should not be specifically included within the GTV, since there is no evidence that recurrence is more likely within the dural tail \citep{rogers2017intermediate, rogers2020high}.
    
    Therefore, future studies that focus on automated segmentation of meningioma for radiation therapy planning should place emphasis on evaluation of the enhancing tumor and post-op resection bed volumes on post-operative T1Gd treatment planning MRI, while reducing emphasis on the surrounding non-enhancing FLAIR hyperintensity compartment and the small trailing linear dural tail enhancement.
    
    However, in the post-treatment follow-up setting, RTOG 0539 still required the use of T1, T2, FLAIR, and T1Gd series, which is similar to the imaging used in the BraTS 2023 Challenge: Intracranial Meningioma. For this challenge’s pre-operative meningioma dataset, the enhancing and non-enhancing tumor labels compose the tumor-core. The SNFH label representing edema, which was included in the WT region of interest is not typically included within radiation therapy meningioma target volumes.  

\section{Conclusion}
    The BraTS 2023 Intracranial Meningioma Segmentation Challenge has marked a significant step forward in the segmentation of meningioma tumors, highlighting both the potential and limitations of current methodologies. As the field moves forward, a focus on enhancing dataset diversity, refining pre-processing techniques, and tailoring segmentation tasks to specific clinical needs will be crucial in translating these advancements into clinical practice.

%%%%%%%%%%%%%%%%%%%%%%%%%%%%%%%%%%%%%%%%%%%%%%%%%%%%%%%%%%%%%%%%%%%%%%%
% Mandatory Sections. Please complete, especially for final publication
%%%%%%%%%%%%%%%%%%%%%%%%%%%%%%%%%%%%%%%%%%%%%%%%%%%%%%%%%%%%%%%%%%%%%%%

% Acknowledgements.
% Please include any funding, intellectual contributions not included in the authorship, and any other acknowledgements.
\acks{Research reported in this publication was partly supported by the National Institutes of Health (NIH) under award numbers: NCI K08CA256045, NCI/ITCR U24CA279629, and NCI/ITCR U01CA242871. The content of this publication is solely the responsibility of the authors and does not represent the official views of the NIH.
        
        Developing large and well curated mpMRI datasets for automated segmentation model development requires significant time and expertise from neuro-radiology experts. We are grateful to everyone who contributed to the development and review of the tumor volume labels including volunteer annotators/approvers from the American Society of Neuroradiology.}

% Ethical Standards.
% Please edit with the appropriate ethics considerations for your work. Include any pertinent IRB information, etc.
%
% Please note that the submission requirements included:
% The work presented must follow appropriate ethical standards in conducting research and writing the manuscript, following all applicable laws and regulations regarding treatment of animals or human subjects.
\ethics{The work follows appropriate ethical standards in conducting research and writing the manuscript, following all applicable laws and regulations regarding treatment of animals and human subjects. All participating sites had institutional review board (IRB) approval. A waiver for informed consent was provided by each institution's respective IRB. }

% Conflict of Interest
% Declaration of possible conflicts of interest: Authors must disclose any financial, organisational, commercial or personal conflicts of interest that might bias their work.
% If no conflicts, please say "We declare we don't have conflicts of interest."
\coi{We declare we do not have conflicts of interest.}

% Data availability
\data{The BraTS Meningioma Pre-operative Dataset training (1,000/1,424, 70\%) and validation (141/1,424, 10\%) data are publicly available on Synapse \citep{calabrese2023brats, LaBella2024}. The testing dataset (283/1,424, 20\%) will be kept private for the foreseeable future to allow for the unbiased assessment of future segmentation algorithms.}

\bibliography{bibliography}

\begin{thebibliography}{51}
\providecommand{\natexlab}[1]{#1}
\providecommand{\url}[1]{\texttt{#1}}
\expandafter\ifx\csname urlstyle\endcsname\relax
  \providecommand{\doi}[1]{doi: #1}\else
  \providecommand{\doi}{doi: \begingroup \urlstyle{rm}\Url}\fi

\bibitem[bra(2023)]{brats_synapse}
Synapse: Brain tumor segmentation (brats) cluster of challenges.
\newblock \url{https://www.synapse.org/#!Synapse:syn51156910/wiki/}, 2023.

\bibitem[Akiba et~al.(2019)Akiba, Sano, Yanase, Ohta, and Koyama]{optuna_2019}
Takuya Akiba, Shotaro Sano, Toshihiko Yanase, Takeru Ohta, and Masanori Koyama.
\newblock Optuna: A next-generation hyperparameter optimization framework.
\newblock In \emph{Proceedings of the 25th {ACM} {SIGKDD} International Conference on Knowledge Discovery and Data Mining}, 2019.

\bibitem[Baid et~al.(2021)Baid, Ghodasara, Mohan, Bilello, Calabrese, Colak, Farahani, Kalpathy-Cramer, Kitamura, Pati, et~al.]{baid2021rsna}
Ujjwal Baid, Satyam Ghodasara, Suyash Mohan, Michel Bilello, Evan Calabrese, Errol Colak, Keyvan Farahani, Jayashree Kalpathy-Cramer, Felipe~C Kitamura, Sarthak Pati, et~al.
\newblock The rsna-asnr-miccai brats 2021 benchmark on brain tumor segmentation and radiogenomic classification.
\newblock \emph{arXiv preprint arXiv:2107.02314}, 2021.

\bibitem[Bakas et~al.(2017)Bakas, Akbari, Sotiras, Bilello, Rozycki, Kirby, Freymann, Farahani, and Davatzikos]{bakas2017advancing}
Spyridon Bakas, Hamed Akbari, Aristeidis Sotiras, Michel Bilello, Martin Rozycki, Justin~S Kirby, John~B Freymann, Keyvan Farahani, and Christos Davatzikos.
\newblock Advancing the cancer genome atlas glioma mri collections with expert segmentation labels and radiomic features.
\newblock \emph{Scientific data}, 4\penalty0 (1):\penalty0 1--13, 2017.

\bibitem[Bischoff-Grethe et~al.(2007)Bischoff-Grethe, Ozyurt, Busa, Quinn, Fennema-Notestine, Clark, Morris, Bondi, Jernigan, Dale, et~al.]{bischoff2007technique}
Amanda Bischoff-Grethe, I~Burak Ozyurt, Evelina Busa, Brian~T Quinn, Christine Fennema-Notestine, Camellia~P Clark, Shaunna Morris, Mark~W Bondi, Terry~L Jernigan, Anders~M Dale, et~al.
\newblock A technique for the deidentification of structural brain mr images.
\newblock \emph{Human brain mapping}, 28\penalty0 (9):\penalty0 892--903, 2007.

\bibitem[Bouget et~al.(2022)Bouget, Pedersen, Jakola, Kavouridis, Emblem, Eijgelaar, Kommers, Ardon, Barkhof, Bello, et~al.]{bouget2022preoperative}
David Bouget, Andr{\'e} Pedersen, Asgeir~S Jakola, Vasileios Kavouridis, Kyrre~E Emblem, Roelant~S Eijgelaar, Ivar Kommers, Hilko Ardon, Frederik Barkhof, Lorenzo Bello, et~al.
\newblock Preoperative brain tumor imaging: Models and software for segmentation and standardized reporting.
\newblock \emph{Frontiers in neurology}, 13:\penalty0 932219, 2022.

\bibitem[Calabrese and LaBella(2023)]{calabrese2023brats}
E.~Calabrese and D.~LaBella.
\newblock 2023 brats meningioma dataset.
\newblock \url{https://www.synapse.org/#!Synapse:syn51514106}, 2023.

\bibitem[Capellan-Martin(2023)]{capellan2023brats}
D.~Capellan-Martin.
\newblock Brats challenge 2023: Model ensemble for brain tumor segmentation in mri.
\newblock In \emph{MICCAI}, Vancouver, Canada, 2023.

\bibitem[Consortium(2020)]{monai2020project}
The~MONAI Consortium.
\newblock Project monai.
\newblock \url{https://doi.org/10.5281/zenodo.4323059}, 2020.

\bibitem[Galldiks et~al.(2023)Galldiks, Albert, Wollring, Werner, Lohmann, Villanueva-Meyer, Fink, Langen, and Tonn]{galldiks2023advances}
Norbert Galldiks, Nathalie~L Albert, Michael Wollring, Jan-Michael Werner, Philipp Lohmann, Javier~E Villanueva-Meyer, Gereon~R Fink, Karl-Josef Langen, and Joerg-Christian Tonn.
\newblock Advances in pet imaging for meningioma patients.
\newblock \emph{Neuro-oncology advances}, 5\penalty0 (Supplement\_1):\penalty0 i84--i93, 2023.

\bibitem[Gilard et~al.(2021)Gilard, Tebani, Dabaj, Laquerri{\`e}re, Fontanilles, Derrey, Marret, and Bekri]{gilard2021diagnosis}
Vianney Gilard, Abdellah Tebani, Ivana Dabaj, Annie Laquerri{\`e}re, Maxime Fontanilles, St{\'e}phane Derrey, St{\'e}phane Marret, and Soumeya Bekri.
\newblock Diagnosis and management of glioblastoma: A comprehensive perspective.
\newblock \emph{Journal of personalized medicine}, 11\penalty0 (4):\penalty0 258, 2021.

\bibitem[Gordillo et~al.(2013)Gordillo, Montseny, and Sobrevilla]{gordillo2013state}
Nelly Gordillo, Eduard Montseny, and Pilar Sobrevilla.
\newblock State of the art survey on mri brain tumor segmentation.
\newblock \emph{Magnetic resonance imaging}, 31\penalty0 (8):\penalty0 1426--1438, 2013.

\bibitem[Havaei et~al.(2017)Havaei, Davy, Warde-Farley, Biard, Courville, Bengio, Pal, Jodoin, and Larochelle]{havaei2017brain}
Mohammad Havaei, Axel Davy, David Warde-Farley, Antoine Biard, Aaron Courville, Yoshua Bengio, Chris Pal, Pierre-Marc Jodoin, and Hugo Larochelle.
\newblock Brain tumor segmentation with deep neural networks.
\newblock \emph{Medical image analysis}, 35:\penalty0 18--31, 2017.

\bibitem[Huang et~al.(2023{\natexlab{a}})Huang, Wang, Deng, Ye, Su, Sun, He, Gu, Gu, and Zhang]{huang2023stu}
Z.~Huang, H.~Wang, Z.~Deng, J.~Ye, Y.~Su, H.~Sun, J.~He, Y.~Gu, L.~Gu, and S.~Zhang.
\newblock Stu-net: Scalable and transferable medical image segmentation models empowered by large-scale supervised pre-training.
\newblock \emph{arXiv preprint arXiv:2304.06716}, 2023{\natexlab{a}}.

\bibitem[Huang et~al.(2023{\natexlab{b}})]{huang2023exploring}
Z.~Huang et~al.
\newblock Exploring model size and patch size for brats23 challenge.
\newblock In \emph{MICCAI}, Vancouver, Canada, 2023{\natexlab{b}}.

\bibitem[Isensee et~al.(2021)Isensee, Jaeger, Kohl, Petersen, and Maier-Hein]{isensee2021nnu}
Fabian Isensee, Paul~F Jaeger, Simon~AA Kohl, Jens Petersen, and Klaus~H Maier-Hein.
\newblock nnu-net: a self-configuring method for deep learning-based biomedical image segmentation.
\newblock \emph{Nature methods}, 18\penalty0 (2):\penalty0 203--211, 2021.

\bibitem[I{\c{s}}{\i}n et~al.(2016)I{\c{s}}{\i}n, Direko{\u{g}}lu, and {\c{S}}ah]{icsin2016review}
Ali I{\c{s}}{\i}n, Cem Direko{\u{g}}lu, and Melike {\c{S}}ah.
\newblock Review of mri-based brain tumor image segmentation using deep learning methods.
\newblock \emph{Procedia Computer Science}, 102:\penalty0 317--324, 2016.

\bibitem[Juluru et~al.(2020)Juluru, Siegel, and Mazura]{juluru2020identification}
Krishna Juluru, Eliot Siegel, and Jan Mazura.
\newblock Identification from mri with face-recognition software.
\newblock \emph{The New England Journal of Medicine}, 382\penalty0 (5):\penalty0 489--490, 2020.

\bibitem[Karargyris et~al.(2023)Karargyris, Umeton, Sheller, Aristizabal, George, Wuest, Pati, Kassem, Zenk, Baid, et~al.]{karargyris2023federated}
Alexandros Karargyris, Renato Umeton, Micah~J Sheller, Alejandro Aristizabal, Johnu George, Anna Wuest, Sarthak Pati, Hasan Kassem, Maximilian Zenk, Ujjwal Baid, et~al.
\newblock Federated benchmarking of medical artificial intelligence with medperf.
\newblock \emph{Nature Machine Intelligence}, 5\penalty0 (7):\penalty0 799--810, 2023.

\bibitem[Kazerooni et~al.(2023)Kazerooni, Khalili, Liu, Haldar, Jiang, Anwar, Albrecht, Adewole, Anazodo, Anderson, et~al.]{kazerooni2023brain}
Anahita~Fathi Kazerooni, Nastaran Khalili, Xinyang Liu, Debanjan Haldar, Zhifan Jiang, Syed~Muhammed Anwar, Jake Albrecht, Maruf Adewole, Udunna Anazodo, Hannah Anderson, et~al.
\newblock The brain tumor segmentation (brats) challenge 2023: Focus on pediatrics (cbtn-connect-dipgr-asnr-miccai brats-peds).
\newblock \emph{ArXiv}, 2023.

\bibitem[Khandwala et~al.(2021)Khandwala, Mubarak, and Minhas]{khandwala2021many}
Kumail Khandwala, Fatima Mubarak, and Khurram Minhas.
\newblock The many faces of glioblastoma: Pictorial review of atypical imaging features.
\newblock \emph{The Neuroradiology Journal}, 34\penalty0 (1):\penalty0 33--41, 2021.

\bibitem[LaBella(2024)]{dlabella29_meningiomaanalysis_2024}
Dominic LaBella.
\newblock {MeningiomaAnalysis}.
\newblock \url{https://doi.org/10.5281/zenodo.13936365}, 2024.

\bibitem[LaBella et~al.(2023)LaBella, Adewole, Alonso-Basanta, Altes, Anwar, Baid, Bergquist, Bhalerao, Chen, Chung, et~al.]{labella2023asnr}
Dominic LaBella, Maruf Adewole, Michelle Alonso-Basanta, Talissa Altes, Syed~Muhammad Anwar, Ujjwal Baid, Timothy Bergquist, Radhika Bhalerao, Sully Chen, Verena Chung, et~al.
\newblock The asnr-miccai brain tumor segmentation (brats) challenge 2023: Intracranial meningioma.
\newblock \emph{arXiv preprint arXiv:2305.07642}, 2023.

\bibitem[LaBella et~al.(2024)LaBella, Khanna, McBurney-Lin, Mclean, Nedelec, Rashid, hoda Tahon, Altes, Baid, Bhalerao, Dhemesh, Floyd, Godfrey, Hilal, Janas, Kazerooni, Kent, Kirkpatrick, Kofler, Leu, Maleki, Menze, Pajot, Reitman, Rudie, Saluja, Velichko, Wang, Warman, Sollmann, Diffley, Nandolia, Warren, Hussain, Fehringer, Bronstein, Deptula, Stein, Taherzadeh, de~Oliveira, Haughey, Kontzialis, Saba, Turner, Brüßeler, Ansari, Gkampenis, Weiss, Mansour, Shawali, Yordanov, Stein, Hourani, Moshebah, Abouelatta, Rizvi, Willms, Martin, Okar, D’Anna, Taha, Sharifi, Faghani, Kite, Pinho, Haider, Alonso-Basanta, Villanueva-Meyer, Rauschecker, Nada, Aboian, Flanders, Bakas, and Calabrese]{LaBella2024}
Dominic LaBella, Omaditya Khanna, Shan McBurney-Lin, Ryan Mclean, Pierre Nedelec, Arif~S. Rashid, Nourel hoda Tahon, Talissa Altes, Ujjwal Baid, Radhika Bhalerao, Yaseen Dhemesh, Scott Floyd, Devon Godfrey, Fathi Hilal, Anastasia Janas, Anahita Kazerooni, Collin Kent, John Kirkpatrick, Florian Kofler, Kevin Leu, Nazanin Maleki, Bjoern Menze, Maxence Pajot, Zachary~J. Reitman, Jeffrey~D. Rudie, Rachit Saluja, Yury Velichko, Chunhao Wang, Pranav~I. Warman, Nico Sollmann, David Diffley, Khanak~K. Nandolia, Daniel~I Warren, Ali Hussain, John~Pascal Fehringer, Yulia Bronstein, Lisa Deptula, Evan~G. Stein, Mahsa Taherzadeh, Eduardo~Portela de~Oliveira, Aoife Haughey, Marinos Kontzialis, Luca Saba, Benjamin Turner, Melanie M.~T. Brüßeler, Shehbaz Ansari, Athanasios Gkampenis, David~Maximilian Weiss, Aya Mansour, Islam~H. Shawali, Nikolay Yordanov, Joel~M. Stein, Roula Hourani, Mohammed~Yahya Moshebah, Ahmed~Magdy Abouelatta, Tanvir Rizvi, Klara Willms, Dann~C. Martin, Abdullah Okar, Gennaro D’Anna, Ahmed Taha,
  Yasaman Sharifi, Shahriar Faghani, Dominic Kite, Marco Pinho, Muhammad~Ammar Haider, Michelle Alonso-Basanta, Javier Villanueva-Meyer, Andreas~M. Rauschecker, Ayman Nada, Mariam Aboian, Adam Flanders, Spyridon Bakas, and Evan Calabrese.
\newblock A multi-institutional meningioma mri dataset for automated multi-sequence image segmentation.
\newblock \emph{Scientific Data}, 11:\penalty0 496, 2024.
\newblock \doi{10.1038/s41597-024-03350-9}.
\newblock URL \url{https://doi.org/10.1038/s41597-024-03350-9}.

\bibitem[Martz et~al.(2022)Martz, Salleron, Dhermain, Vogin, Daisne, Audouard, Tanguy, Noel, Peyre, Lecouillard, et~al.]{martz2022anocef}
N~Martz, J~Salleron, F~Dhermain, G~Vogin, JF~Daisne, R~Mouttet Audouard, R~Tanguy, G~Noel, M~Peyre, I~Lecouillard, et~al.
\newblock Anocef consensus guideline on target volume delineation for meningiomas radiotherapy.
\newblock \emph{International Journal of Radiation Oncology, Biology, Physics}, 114\penalty0 (3):\penalty0 e46, 2022.

\bibitem[Menze et~al.(2014)Menze, Jakab, Bauer, Kalpathy-Cramer, Farahani, Kirby, Burren, Porz, Slotboom, Wiest, et~al.]{menze2014multimodal}
Bjoern~H Menze, Andras Jakab, Stefan Bauer, Jayashree Kalpathy-Cramer, Keyvan Farahani, Justin Kirby, Yuliya Burren, Nicole Porz, Johannes Slotboom, Roland Wiest, et~al.
\newblock The multimodal brain tumor image segmentation benchmark (brats).
\newblock \emph{IEEE transactions on medical imaging}, 34\penalty0 (10):\penalty0 1993--2024, 2014.

\bibitem[{MLCommons Association}(2024)]{MLCube2024}
{MLCommons Association}.
\newblock Mlcube: Standardizing ml deployment.
\newblock \url{https://mlcommons.org/working-groups/data/mlcube/}, 2024.
\newblock Accessed: 2024-05-08.

\bibitem[Moawad et~al.(2023)Moawad, Janas, Baid, Ramakrishnan, Jekel, Krantchev, Moy, Saluja, Osenberg, Wilms, et~al.]{moawad2023brain}
Ahmed~W Moawad, Anastasia Janas, Ujjwal Baid, Divya Ramakrishnan, Leon Jekel, Kiril Krantchev, Harrison Moy, Rachit Saluja, Klara Osenberg, Klara Wilms, et~al.
\newblock The brain tumor segmentation (brats-mets) challenge 2023: Brain metastasis segmentation on pre-treatment mri.
\newblock \emph{ArXiv}, 2023.

\bibitem[Murek(2024)]{murek_localization_2024}
Michael Murek.
\newblock Localization of intracranial meningiomas, 2024.
\newblock URL \url{https://neurochirurgie.insel.ch/en/what-we-treat/brain-tumor/meningioma}.
\newblock Image: University Department of Neurosurgery, Inselspital Bern © CC BY-NC 4.0. Left: Axial view of meningiomas at the base of the skull. Right: Coronal view of meningiomas with location at the cranial dome, the falx cerebri as well as intraventricular.

\bibitem[Myronenko et~al.(2023)Myronenko, Yang, He, and Xu]{myronenko2023auto3dseg}
A.~Myronenko, D.~Yang, Y.~He, and D.~Xu.
\newblock Auto3dseg for brain tumor segmentation from 3d mri in brats 2023 challenge.
\newblock In \emph{MICCAI}, Vancouver, Canada, 2023.

\bibitem[Myronenko(2018)]{myronenko2018mri}
Andriy Myronenko.
\newblock 3d mri brain tumor segmentation using autoencoder regularization.
\newblock In \emph{International MICCAI Brainlesion Workshop}, pages 311--320. Springer, Cham, 2018.

\bibitem[Ogasawara et~al.(2021)Ogasawara, Philbrick, and Adamson]{ogasawara2021meningioma}
Christian Ogasawara, Brandon~D Philbrick, and D~Cory Adamson.
\newblock Meningioma: a review of epidemiology, pathology, diagnosis, treatment, and future directions.
\newblock \emph{Biomedicines}, 9\penalty0 (3):\penalty0 319, 2021.

\bibitem[Pati et~al.(2022)Pati, Baid, Edwards, Sheller, Foley, Reina, Thakur, Sako, Bilello, Davatzikos, et~al.]{pati2022federated}
Sarthak Pati, Ujjwal Baid, Brandon Edwards, Micah~J Sheller, Patrick Foley, G~Anthony Reina, Siddhesh Thakur, Chiharu Sako, Michel Bilello, Christos Davatzikos, et~al.
\newblock The federated tumor segmentation (fets) tool: an open-source solution to further solid tumor research.
\newblock \emph{Physics in Medicine \& Biology}, 67\penalty0 (20):\penalty0 204002, 2022.

\bibitem[Pati et~al.(2023)Pati, Thakur, Hamamc{\i}, Baid, Baheti, Bhalerao, G{\"u}ley, Mouchtaris, Lang, Thermos, et~al.]{pati2023gandlf}
Sarthak Pati, Siddhesh~P Thakur, {\.I}brahim~Ethem Hamamc{\i}, Ujjwal Baid, Bhakti Baheti, Megh Bhalerao, Orhun G{\"u}ley, Sofia Mouchtaris, David Lang, Spyridon Thermos, et~al.
\newblock Gandlf: the generally nuanced deep learning framework for scalable end-to-end clinical workflows.
\newblock \emph{Communications Engineering}, 2\penalty0 (1):\penalty0 23, 2023.

\bibitem[Pereira et~al.(2016)Pereira, Pinto, Alves, and Silva]{pereira2016brain}
S{\'e}rgio Pereira, Adriano Pinto, Victor Alves, and Carlos~A Silva.
\newblock Brain tumor segmentation using convolutional neural networks in mri images.
\newblock \emph{IEEE transactions on medical imaging}, 35\penalty0 (5):\penalty0 1240--1251, 2016.

\bibitem[Prasad et~al.(2022)Prasad, Perlow, Bovi, Braunstein, Ivanidze, Kalapurakal, Kleefisch, Knisely, Mehta, Prevedello, et~al.]{prasad202268ga}
Rahul~N Prasad, Haley~K Perlow, Joseph Bovi, Steve~E Braunstein, Jana Ivanidze, John~A Kalapurakal, Christopher Kleefisch, Jonathan~PS Knisely, Minesh~P Mehta, Daniel~M Prevedello, et~al.
\newblock 68ga-dotatate pet: the future of meningioma treatment.
\newblock \emph{International Journal of Radiation Oncology, Biology, Physics}, 113\penalty0 (4):\penalty0 868--871, 2022.

\bibitem[Rogers et~al.(2020)Rogers, Won, Vogelbaum, Perry, Ashby, Modi, Alleman, Galvin, Fogh, Youssef, et~al.]{rogers2020high}
C~Leland Rogers, Minhee Won, Michael~A Vogelbaum, Arie Perry, Lynn~S Ashby, Jignesh~M Modi, Anthony~M Alleman, James Galvin, Shannon~E Fogh, Emad Youssef, et~al.
\newblock High-risk meningioma: initial outcomes from nrg oncology/rtog 0539.
\newblock \emph{International Journal of Radiation Oncology* Biology* Physics}, 106\penalty0 (4):\penalty0 790--799, 2020.

\bibitem[Rogers et~al.(2017)Rogers, Zhang, Vogelbaum, Perry, Ashby, Modi, Alleman, Galvin, Brachman, Jenrette, et~al.]{rogers2017intermediate}
Leland Rogers, Peixin Zhang, Michael~A Vogelbaum, Arie Perry, Lynn~S Ashby, Jignesh~M Modi, Anthony~M Alleman, James Galvin, David Brachman, Joseph~M Jenrette, et~al.
\newblock Intermediate-risk meningioma: initial outcomes from nrg oncology rtog 0539.
\newblock \emph{Journal of neurosurgery}, 129\penalty0 (1):\penalty0 35--47, 2017.

\bibitem[Rudie(2023)]{rudie2023brats}
J.~Rudie.
\newblock Brats 2023 segmentation metrics: Clinical relevance.
\newblock In \emph{Proceedings of the Medical Image Computing and Computer Assisted Intervention -- MICCAI}, Vancouver, Canada, 2023.

\bibitem[Salah et~al.(2019)Salah, Tabbarah, Asmar, Tamim, Makki, Sibahi, Hourani, et~al.]{salah2019can}
Florian Salah, A~Tabbarah, K~Asmar, H~Tamim, M~Makki, A~Sibahi, R~Hourani, et~al.
\newblock Can ct and mri features differentiate benign from malignant meningiomas?
\newblock \emph{Clinical Radiology}, 74\penalty0 (11):\penalty0 898--e15, 2019.

\bibitem[Saluja(2023)]{saluja2023lesion}
R.~Saluja.
\newblock Lesion-wise performance metrics for brats-2023 segmentation challenges.
\newblock In \emph{Proceedings of the Medical Image Computing and Computer Assisted Intervention -- MICCAI}, Vancouver, Canada, 2023.

\bibitem[Schwarz et~al.(2019)Schwarz, Kremers, Therneau, Sharp, Gunter, Vemuri, Arani, Spychalla, Kantarci, Knopman, et~al.]{schwarz2019identification}
Christopher~G Schwarz, Walter~K Kremers, Terry~M Therneau, Richard~R Sharp, Jeffrey~L Gunter, Prashanthi Vemuri, Arvin Arani, Anthony~J Spychalla, Kejal Kantarci, David~S Knopman, et~al.
\newblock Identification of anonymous mri research participants with face-recognition software.
\newblock \emph{New England Journal of Medicine}, 381\penalty0 (17):\penalty0 1684--1686, 2019.

\bibitem[Schwarz et~al.(2021)Schwarz, Kremers, Wiste, Gunter, Vemuri, Spychalla, Kantarci, Schultz, Sperling, Knopman, et~al.]{schwarz2021changing}
Christopher~G Schwarz, Walter~K Kremers, Heather~J Wiste, Jeffrey~L Gunter, Prashanthi Vemuri, Anthony~J Spychalla, Kejal Kantarci, Aaron~P Schultz, Reisa~A Sperling, David~S Knopman, et~al.
\newblock Changing the face of neuroimaging research: comparing a new mri de-facing technique with popular alternatives.
\newblock \emph{NeuroImage}, 231:\penalty0 117845, 2021.

\bibitem[Smith(2002)]{smith2002fast}
Stephen~M Smith.
\newblock Fast robust automated brain extraction.
\newblock \emph{Human brain mapping}, 17\penalty0 (3):\penalty0 143--155, 2002.

\bibitem[Tang et~al.(2022)Tang, Yang, Li, Roth, et~al.]{tang2022self}
Yucheng Tang, Dong Yang, Wenqi Li, Holger~R. Roth, et~al.
\newblock Self-supervised pre-training of swin transformers for 3d medical image analysis.
\newblock In \emph{Proceedings of the IEEE/CVF Conference on Computer Vision and Pattern Recognition (CVPR)}, pages 20730--20740, 2022.

\bibitem[Thakur et~al.(2020)Thakur, Doshi, Pati, Rathore, Sako, Bilello, Ha, Shukla, Flanders, Kotrotsou, et~al.]{thakur2020brain}
Siddhesh Thakur, Jimit Doshi, Sarthak Pati, Saima Rathore, Chiharu Sako, Michel Bilello, Sung~Min Ha, Gaurav Shukla, Adam Flanders, Aikaterini Kotrotsou, et~al.
\newblock Brain extraction on mri scans in presence of diffuse glioma: Multi-institutional performance evaluation of deep learning methods and robust modality-agnostic training.
\newblock \emph{NeuroImage}, 220:\penalty0 117081, 2020.

\bibitem[Thakur et~al.(2019)Thakur, Doshi, Pati, Ha, Sako, Talbar, Kulkarni, Davatzikos, Erus, and Bakas]{thakur2019skull}
Siddhesh~P Thakur, Jimit Doshi, Sarthak Pati, Sung~Min Ha, Chiharu Sako, Sanjay Talbar, Uday Kulkarni, Christos Davatzikos, Guray Erus, and Spyridon Bakas.
\newblock Skull-stripping of glioblastoma mri scans using 3d deep learning.
\newblock In \emph{International MICCAI Brainlesion Workshop}, pages 57--68. Springer, 2019.

\bibitem[Theyers et~al.(2021)Theyers, Zamyadi, O'Reilly, Bartha, Symons, MacQueen, Hassel, Lerch, Anagnostou, Lam, et~al.]{theyers2021multisite}
Athena~E Theyers, Mojdeh Zamyadi, Mark O'Reilly, Robert Bartha, Sean Symons, Glenda~M MacQueen, Stefanie Hassel, Jason~P Lerch, Evdokia Anagnostou, Raymond~W Lam, et~al.
\newblock Multisite comparison of mri defacing software across multiple cohorts.
\newblock \emph{Frontiers in psychiatry}, 12:\penalty0 617997, 2021.

\bibitem[Wasserthal et~al.(2022)Wasserthal, Meyer, Breit, Cyriac, Yang, and Segeroth]{wasserthal2022totalsegmentator}
J.~Wasserthal, M.~Meyer, H.C. Breit, J.~Cyriac, S.~Yang, and M.~Segeroth.
\newblock Totalsegmentator: robust segmentation of 104 anatomical structures in ct images.
\newblock \emph{arXiv preprint arXiv:2208.05868}, 2022.

\bibitem[Watts et~al.(2014)Watts, Box, Galvin, Brotchie, Trost, and Sutherland]{watts2014magnetic}
J~Watts, G~Box, A~Galvin, P~Brotchie, N~Trost, and T~Sutherland.
\newblock Magnetic resonance imaging of meningiomas: a pictorial review.
\newblock \emph{Insights into imaging}, 5\penalty0 (1):\penalty0 113--122, 2014.

\bibitem[Yushkevich et~al.(2006)Yushkevich, Piven, Cody~Hazlett, Gimpel~Smith, Ho, Gee, and Gerig]{py06nimg}
Paul~A. Yushkevich, Joseph Piven, Heather Cody~Hazlett, Rachel Gimpel~Smith, Sean Ho, James~C. Gee, and Guido Gerig.
\newblock User-guided {3D} active contour segmentation of anatomical structures: Significantly improved efficiency and reliability.
\newblock \emph{Neuroimage}, 31\penalty0 (3):\penalty0 1116--1128, 2006.

\end{thebibliography}

% needed in general before appendices.
% \newpage

\begin{comment}

% Appendix is optional
\clearpage
\appendix
\section{Proof of the central theorem}
	In this appendix we prove the central theorem and present additional experimental results.
	\noindent

	{\noindent \em Remainder omitted in this sample. }

\section{Appendix section}
	\subsection{Appendix subsection}
		\subsubsection{Appendix subsubsection}
			\paragraph{Appendix paragraph} Lorem ipsum dolor sit amet, consectetur adipisicing elit, sed do eiusmod
			tempor incididunt ut labore et dolore magna aliqua. Ut enim ad minim veniam,
			quis nostrud exercitation ullamco laboris nisi ut aliquip ex ea commodo
			consequat. Duis aute irure dolor in reprehenderit in voluptate velit esse
			cillum dolore eu fugiat nulla pariatur. Excepteur sint occaecat cupidatat non
			proident, sunt in culpa qui officia deserunt mollit anim id est laborum.
    
\end{comment}
\end{document}